\documentclass[11pt,a4paper]{article}

\usepackage{microtype}
\usepackage[linesnumbered,vlined]{algorithm2e}
\usepackage{amsmath, amssymb, enumerate}
\usepackage{breqn}
\usepackage{authblk}
\usepackage{adjustbox}
\usepackage{subcaption}
\usepackage{hyperref}
\usepackage[margin=1.1in]{geometry}
\def\fixme#1{\typeout{FIXED in page \thepage : {#1}}
\bgroup \color{red}{[FIXME: {#1}]} \egroup}

\title{Deterministic Memory Abstraction and Supporting Multicore System
  Architecture\footnote{This paper is accepted to appear in ECRTS 2018.}}
\date{}

\begin{document}

\author{Farzad~Farshchi$^\dagger$, Prathap Kumar Valsan$^\star$,
Renato Mancuso$^\ddagger$, 
Heechul Yun$^\dagger$ \\
$^\dagger$ University of Kansas,  \{farshchi, heechul.yun\}@ku.edu\\
$^\star$ Intel, prathap.kumar.valsan@intel.com\\
$^\ddagger$ Boston University,
rmancuso@bu.edu \\
}

\maketitle

\begin{abstract}
Poor time predictability of multicore processors has been a long-standing
challenge in the real-time systems community.
In this paper, we make a
case that a fundamental problem that prevents efficient and
predictable real-time computing on multicore is the \emph{lack of a proper
memory abstraction} to express memory criticality, which cuts across
various layers of the system: the application, OS, and hardware.
We, therefore, propose a new holistic resource management approach
driven by a new memory abstraction, which we call \emph{Deterministic
 Memory}. The key characteristic of deterministic memory is that the
platform---the OS and hardware---guarantees small and tightly bounded
worst-case memory access timing. In contrast, we call the conventional
memory abstraction as best-effort memory in which only highly
pessimistic worst-case bounds can be achieved. We propose to utilize
both abstractions to achieve high time predictability but
without significantly sacrificing performance.
We present deterministic memory-aware OS and architecture
designs, including OS-level page allocator, hardware-level cache, and
DRAM controller designs. We implement the proposed OS and
architecture extensions on Linux and gem5 simulator. Our evaluation
results, using a set of synthetic and real-world benchmarks,
demonstrate the feasibility and effectiveness of our approach.

 \end{abstract}

\section{Introduction} \label{sec:intro}
High-performance embedded multicore platforms are increasingly
demanded in cyber-physical systems (CPS)---especially those in
automotive and aviation applications---to cut cost and to reduce size,
weight, and power (SWaP) of the system via
consolidation~\cite{ospert2015keynote}. 

Consolidating multiple tasks with different criticality levels
(a.k.a. mixed-criticality
systems~\cite{vestal2007preemptive,burns2013mixed}) on a single
multicore processor is, however, extremely challenging because
interference in shared hardware resources in the memory hierarchy
can significantly alter the tasks' timing characteristics. Poor time
predictability of multicore platforms is a major hurdle that makes
their adoption challenging in many safety-critical CPS. For example,
the CAST-32A position paper by the 
avionics certification authorities comprehensively discusses the
certification challenges of multicore avionics~\cite{faa2016certification}.
Therefore, in the aerospace industry, it is a common practice to disable
all but one core~\cite{kotaba2013multicore},
because extremely pessimistic worst-case-execution times (WCETs)
nullify any performance benefits of using multicore processors in
critical applications. This phenomenon is also known as the
``one-out-of-m'' problem~\cite{kim2016attacking}.


There have been significant
research efforts to address the problem. Two common strategies are (1)
partitioning the shared resources among the tasks or cores to achieve
spatial isolation and (2) applying analyzable arbitration schemes
(e.g., time-division multiple access) in accessing the shared resources to achieve
temporal isolation. These strategies have been studied individually (e.g.,
cache~\cite{kim2013coordinated,ward2013ecrts,mancuso2013rtas}, DRAM
banks~\cite{liu2012software,yun2014rtas}, memory 
bus~\cite{yun2013rtas,nowotsch2014multi}) or in 
combination (e.g., ~\cite{kim2016attacking,suzuki2013coordinated}).
However, most of these efforts improve predictability at the cost of
a significant \emph{sacrifice in  efficiency and performance}.

In this paper, we argue that the fundamental problem that prevents
efficient and predictable real-time computing on multicore is the
\emph{lack of a proper memory abstraction} to express memory
criticality, which cuts across various layers of the system: the
application, OS, and hardware.
Thus, our approach starts by defining a new OS-level
memory abstraction, which we call \emph{Deterministic Memory}.
The key characteristic of deterministic memory is that the
platform---the OS and hardware---guarantees small and tightly bounded
worst-case memory access timing. In contrast, we call the conventional
memory abstraction as best-effort memory in which only highly
pessimistic worst-case bounds can be achieved.


We propose a new holistic cross-layer resource management approach that
leverages the deterministic and best-effort memory abstractions. In
our approach, a task can allocate either type of memory blocks in its
address space, at the page granularity, based on the desired
WCET requirement in accessing the memory blocks. The OS and hardware
then apply different resource management strategies depending on the
memory type. Specifically, predictability focused strategies, such as
resource reservation and predictable scheduling, shall be used for
deterministic memory while average performance and efficiency-focused
strategies, such as space sharing and out-of-order scheduling, shall
be used for best-effort memory.
Because neither all tasks are time-critical nor all memory blocks of a 
time-critical task are  equally important with respect to the task's
WCET, our approach enables the possibility of achieving high time
predictability without significantly affecting performance and
efficiency through the selective use of deterministic memory.




While our approach is a generic framework that can be applied to any
shared hardware resource management, in this paper, we particularly
focus on the shared cache and main memory, and demonstrate the potential
benefits of our approach in the context of shared cache and DRAM
related resource management.
First, we describe OS extensions and an OS-level memory
allocation method to support deterministic memory.
We then describe a deterministic memory-aware cache
design that provides the same level of cache space isolation of the
conventional way-based partitioning techniques, while achieving
significantly higher cache space utilization.
We also describe a deterministic memory-aware DRAM controller design
that extends a previously proposed real-time memory
controller~\cite{valsan2015cpsna} to achieve similar
predictability benefits with minimal DRAM space waste.


We implement the deterministic memory abstraction and an OS-level
memory allocator (replacing Linux's buddy allocator) in a Linux 3.13  
kernel and implement the proposed deterministic-memory aware memory
hierarchy hardware extensions (in MMU, TLB, cache and DRAM controller)
in a gem5 full-system simulator~\cite{binkert2011gem5} modeling
a high-performance (out-of-order) quad-core platform as the baseline.
We evaluate the system using a set of synthetic and real-world 
benchmarks from EEMBC~\cite{eembc},
SD-VBS~\cite{venkata2009sd} and SPEC2006~\cite{henning2006spec}
suites.
We achieve the same degree of isolation with 
conventional way-based cache partitioning for real-time
tasks while improving the cache hit rate of co-scheduled
non-real-time workloads by 39\% on average.
In addition, we need significantly less memory space in reserved DRAM
banks, while achieving comparable WCET guarantees compared with a
state-of-the-art real-time DRAM controller.

The main contributions of this work are as follows:
\begin{itemize}
\item We propose a new OS-level memory abstraction, which we call
  \emph{Deterministic Memory}, that enables efficient cross-layer
  resource management, balancing time predictability and resource
  efficiency.
\item We present a concrete system design---from the OS down to the
  entire memory hierarchy, including shared cache and DRAM controller
  designs---that demonstrate the potential benefits of the new memory
  abstraction. The key contribution of our design is its Memory
  Management Unit (MMU) based approach that provides flexible,
  fine-grained (page-granularity) resource management across the
  entire memory hierarchy.
\item We implement a realistic prototype system on a Linux kernel and
  a cycle-accurate full system simulator.~\footnote{We provide the
    modified Linux kernel source, the modified gem5 simulator source, and
    the simulation methodology at \url{http://github.com/CSL-KU/detmem} for
    replication study.} We also provide extensive empirical results,
  using both synthetic and real-world benchmarks, that demonstrates
  the effectiveness of our approach.
\end{itemize}
The remainder of the paper is organized as
follows. Section~\ref{sec:background} provides
background and motivation. 
Section~\ref{sec:detmem} describes the
proposed Deterministic Memory abstraction.
Section~\ref{sec:design} provides an overview of the deterministic
memory-aware system design.
Section~\ref{sec:analysis} presents DM-aware timing analysis.
Section~\ref{sec:impl} details our prototype implementation.
Section~\ref{sec:evaluation} presents evaluation results. 
We review related work in Section~\ref{sec:related} and
conclude in Section~\ref{sec:conclusion}.

\section{Background and Motivation}\label{sec:background}

In this section, we describe why the standard uniform memory
abstraction is a fundamental limitation for the development of
efficient and predictable real-time computing infrastructures.



{\bf CPU-centric Abstractions and Resource Management.}
Traditionally, the CPU has been the main focus of resource management
in real-time systems. This is because, in a unicore processor, only
one task at a time can access the entire memory hierarchy and that CPU
scheduling decisions have a predominant impact on the response time of
real-time tasks. Therefore, CPU-centric abstractions such as core,
task and task priority have been the primary focus of resource
management. However, in multicore platforms, which have become
mainstream over the last decade, extensive inter-core hardware
resource sharing in the memory hierarchy heavily impacts task timing.
Hence, CPU time management is no longer the sole dimension to explore
when reasoning about the temporal behavior of a system. Various OS and
hardware-level solutions have been proposed to manage shared resources 
in the memory hierarchy with the goal of improving time predictability
(we will provide a comprehensive review of related work in
Section~\ref{sec:related}). Nonetheless, in most approaches,
CPU-centric abstractions are still most widely used to perform
allocation and scheduling of shared resources in the memory
hierarchy. Unfortunately, CPU-centric abstractions are often too
coarse-grained to enact accurate management policies on memory
hierarchy resources, such as cache lines and main memory pages. For
instance, when a fraction of cache space is reserved for a task, it
cannot be used by other tasks, even if it is not fully utilized by the
reserved task.  Likewise, when DRAM banks are reserved for a task,
they cannot be utilized by other tasks, resulting in under-utilized
DRAM space, even though not all memory of the task may need to be
allocated on the reserved DRAM bank.


{\bf The Uniform Memory Abstraction.} Operating systems and hardware
traditionally have provided a simple uniform memory abstraction that
hides all the complex details of the memory hierarchy.  When an
application requests to allocate more memory, the OS simply maps the
necessary amount of \emph{any} physical memory pages available at the
time to the application's address space---without considering: (1) how
the memory pages are actually mapped to the shared hardware resources
in the memory hierarchy, and (2) how they will affect application
performance. Likewise, the underlying hardware 
components treat all memory accesses from the CPU as equal without any
regard to differences in criticality and timing requirements in
allocating and scheduling the requests.

We argue that this uniform memory abstraction is fundamentally
inadequate for multicore systems because it prevents the OS and the
memory hierarchy hardware from making informed decisions in allocating
and scheduling access to shared hardware resources.
As such, we believe that new memory abstractions are needed to enable
\emph{both} efficient and predictable real-time resource management. It is
important to note 
that the said abstractions should not expose too many architectural
details about the memory hierarchy to the users, to ensure portability
in spite of rapid changes in hardware architectures.

\section{Deterministic Memory Abstraction} \label{sec:detmem}

In this section, we introduce the \emph{Deterministic Memory}
abstraction to address the aforementioned challenges.

We define deterministic memory as special memory space for which the
OS and hardware guarantee small and tightly bounded worst-case access
delay. In contrast, we call conventional memory as best-effort memory,
for which only highly pessimistic worst-case bounds can be achieved.
A platform shall support both memory types, which allow applications to
express their memory access timing requirements in an
architecture-neutral way, while leaving the implementation details to
the platform---the OS and the hardware architecture.
This, in turn, enables efficient and analyzable cross-layer resource
management, as we will discuss in the rest of the section.

\begin{figure}[h]
  \centering
  \includegraphics[width=0.7\textwidth]{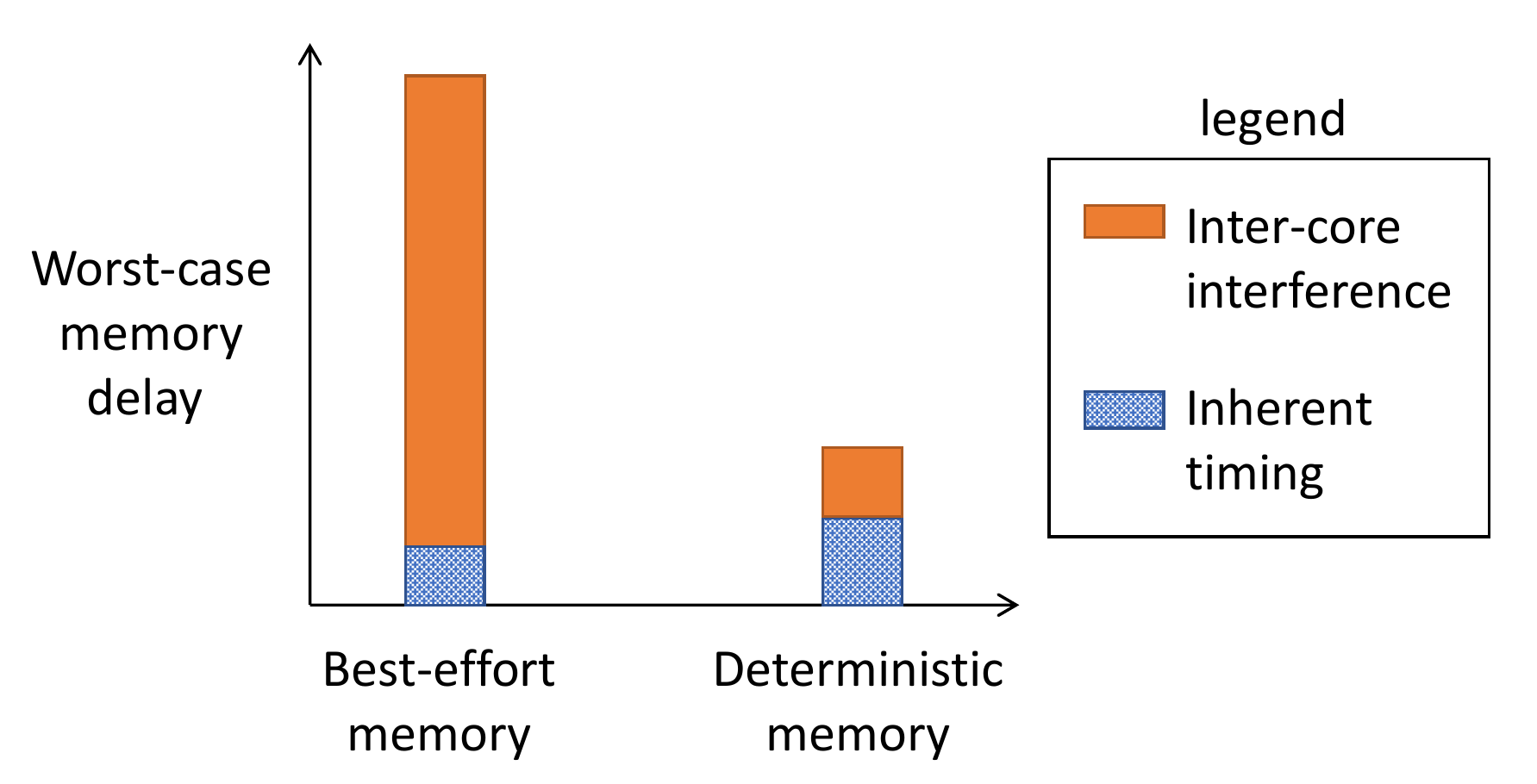}
  \caption{Conceptual differences of deterministic and best-effort memories}
  \label{fig:concept}
\end{figure}

Figure~\ref{fig:concept} shows the conceptual differences between the
two memory types with respect to worst-case memory access delay bounds.
For clarity, we divide memory access delay into two
components: \emph{inherent access delay} and \emph{inter-core interference
  delay}. The inherent access delay is the minimum 
necessary timing in isolation. In this regard, deterministic memory
can be slower---in principal, but not necessarily---than best-effort
memory, as its main objective is predictability and not performance,
while in the case of best-effort memory, the reverse is true.
The inter-core interference delay is, on the other hand, an additional
delay caused by concurrently sharing hardware resources between
multiple cores. This is where the two memory types differ the most.
For best-effort memory, the worst-case delay bound is highly
\emph{pessimistic} mainly because the inter-core interference delay
can be severe.  For deterministic memory, on the other hand, the
worst-case delay bound is \emph{small and tight} as the inter-core
interference delay is minimized by the platform.

\begin{table}[h]
\begin{center}
\begin{tabular}{ r || c | c || c }
  \hline			
                  & Space allocation     & Request scheduling  &  WCET bounds\\
  \hline
  Deterministic  memory & Dedicated resources  & Predictability  focused & Tight \\
  \hline
  Best-effort memory   & Shared  resources & Performance focused  & Pessimistic \\
  \hline  
\end{tabular}
\caption{Differences in resource management strategies.}
\label{tbl:properties}
\end{center}
\end{table}

Table~\ref{tbl:properties} shows general spatial and temporal resource
management strategies of the OS and hardware to achieve the differing
goals of the two memory types.
Here, we mainly focus on shared hardware resources, such as shared
cache, DRAM banks, memory controllers, and buses. In contrast, we
do not focus on core-private hardware resources such as private (L1)
caches as they do not generally contribute to inter-core
interference.



In the deterministic memory approach, a task can map all or part of
its memory from the deterministic memory. For example, an entire
address space of a real-time task can be allocated from the deterministic
memory; or, only the important buffers used in a control loop of the
real-time task can be allocated from the deterministic memory, while
temporary buffers used in the initialization phase are allocated from
the best-effort memory.

Our key insight is that \emph{not all memory blocks
  of an application are equally important with respect to the
  application's WCET.}
For instance, in the applications we profiled in
Section~\ref{sec:profile}, only a small fraction of memory pages
account for most memory accesses to the shared memory
hierarchy (shared cache and DRAM).

Based on this insight, we now provide a detailed design and
implementation of deterministic memory-aware OS and architecture
extensions with a goal of achieving high efficiency and
predictability.


\section{System Design}\label{sec:design}

In this section, we first provide a high-level overview of a
deterministic-memory based multicore system
design (Section~\ref{sec:dm-overview}). We then describe
necessary small OS and hardware architecture extensions to support the
deterministic/best-effort memory abstractions
(Section~\ref{sec:dm-os_arch}). Lastly, we describe deterministic
memory-aware cache and DRAM management 
frameworks (Section~\ref{sec:dm-cache} and \ref{sec:dm-dram},
respectively).

\subsection{Overview}\label{sec:dm-overview}

\begin{figure} [h]
    \centering
    \begin{subfigure}{0.4\textwidth}
        \includegraphics[width=\linewidth]{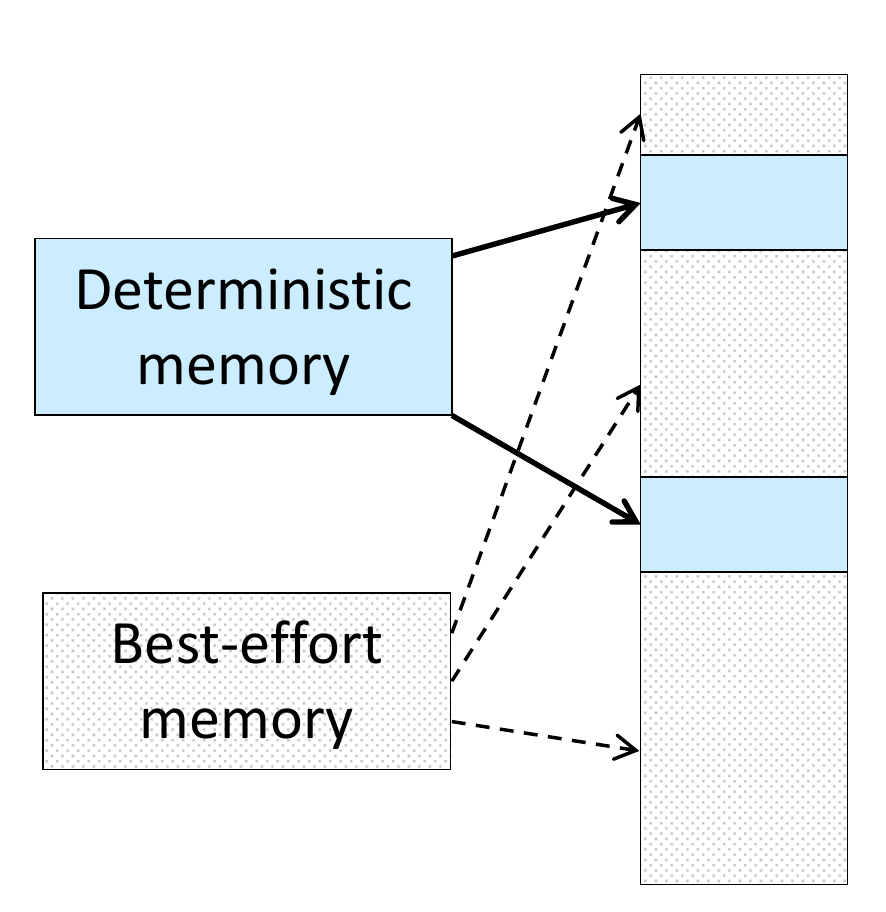}
        \caption{Application view (logical)}
        \label{fig:appview}
    \end{subfigure}
    \hfill
    \begin{subfigure}{0.5\textwidth}
        \includegraphics[width=\linewidth]{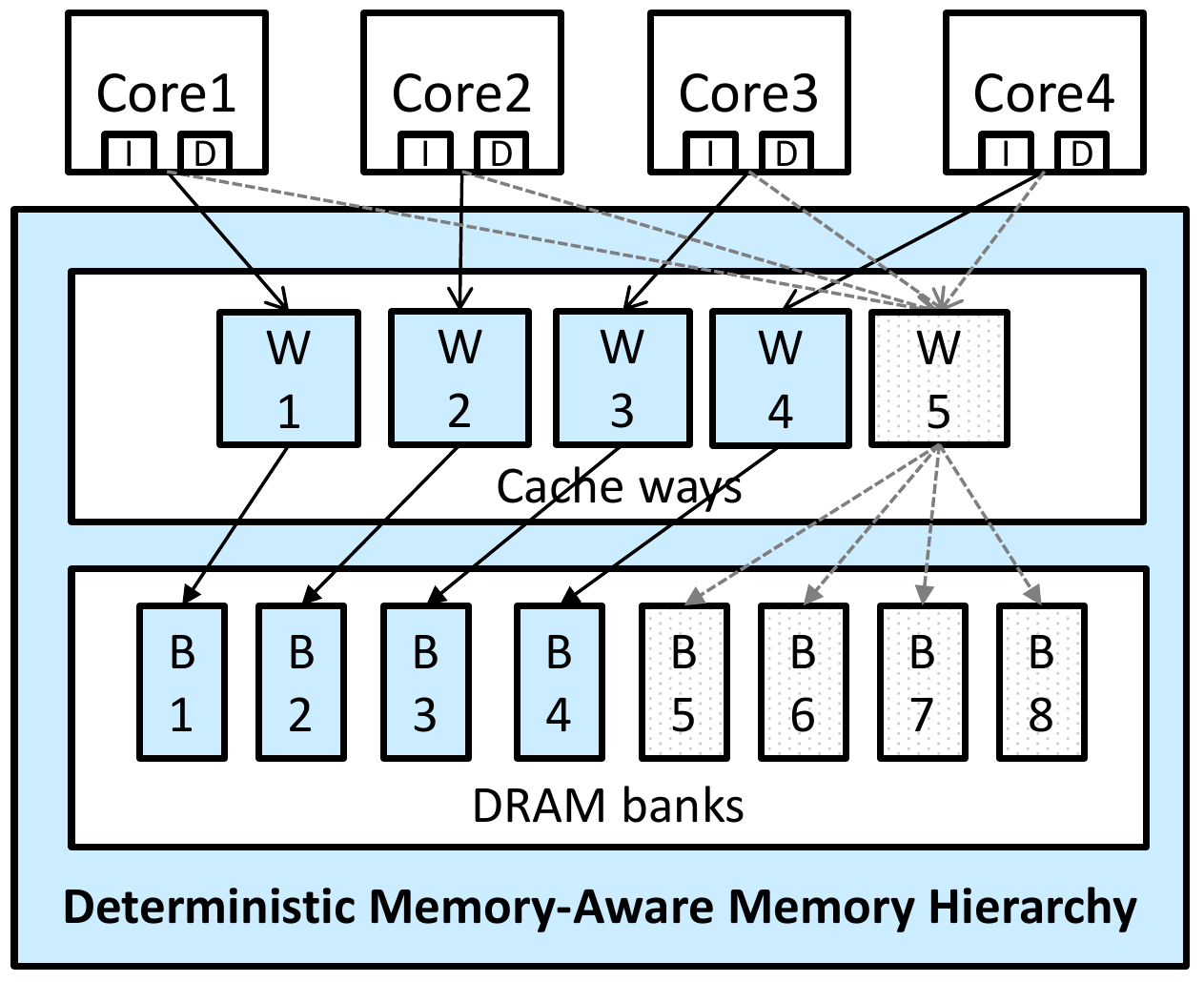}
        \caption{System-level view (physical)}
        \label{fig:osview}
    \end{subfigure}
    \caption{Logical and physical mappings of the \emph{deterministic}
      and \emph{best-effort} memory  abstractions}
    \label{fig:overview}
\end{figure}

Figure~\ref{fig:appview} shows the virtual address space of a task
using both deterministic and best-effort memory under our
approach. From the point of view of the task, the two memory types
differ only in their worst-case timing characteristics. 
The deterministic memory is realized by extending the \emph{virtual
  memory system} at the page granularity. Whether a certain page is
deterministic or best-effort memory is stored in the task's page table
and the information is propagated throughout the shared memory
hierarchy, which is then used in allocation and scheduling decisions
made by the OS and the memory hierarchy hardware.

Figure~\ref{fig:osview} shows the system-level (OS and architecture)
view of a multicore system supporting the deterministic
and best-effort memory abstractions. In this example, each core is
given one cache way and a DRAM bank which will be used to serve
deterministic memory for the core. One cache way and four DRAM banks
are assigned to the best-effort memory of all cores. Here, the
highlighted deterministic memory-aware memory hierarchy refers to
hardware support for the deterministic memory abstraction.

It is important to note that the support for deterministic memory is 
generally more \emph{expensive} than that of best-effort memory in
the sense that it may require dedicated space, which may be wasted if
under-utilized, and predictability focused scheduling, which may not
offer the highest performance. As such, to improve efficiency and
performance, it is desirable to use as little deterministic memory as
possible as long as the desired worst-case timing of real-time tasks
can be satisfied. 

\subsection{OS and Architecture Extensions for Deterministic Memory
  Abstraction Support} \label{sec:dm-os_arch}

The deterministic memory abstraction is realized by extending the OS's
virtual memory subsystem. Whether a certain page has the deterministic
memory property or not is stored in the corresponding page table entry. 
Note that in most architectures, a page table entry contains
not only the virtual-to-physical address translation but also a number
of auxiliary attributes such as access permission and cacheability. The
deterministic memory can be encoded as just another attribute, which
we call a $DM$ bit, in the page table entry.~\footnote{In our
  implementation, we currently use
  an unused memory attribute in the page table entry of the ARM
  architecture; see Section~\ref{sec:impl} for details.}
The OS is responsible for updating the DM bits in each task's page
tables. The OS provides interfaces for applications to declare and
update their deterministic/best-effort memory regions at the page
granularity. Any number of memory regions of any sizes (down to a
single page) within the application's address space can be declared as
deterministic memory (the rest is best-effort memory by default).


In a modern processor, the processor's view of memory is determined by
the Memory Management Unit (MMU), which translates a virtual address
to a corresponding physical address. The translation information,
along with other auxiliary information, is stored in a page table
entry, which is managed by the OS. Translation Look-aside Buffer (TLB)
then caches frequently accessed page table entries to accelerate the
translation speed. As discussed above, in our design, the $DM$ bit in
each page table entry indicates whether the page is for deterministic
memory or for best-effort memory. Thus, the TLB also stores the $DM$
bit and passes the information down to the memory hierarchy.

\begin{figure}[h]
\centering
  \centering
  \includegraphics[width=0.9\textwidth]{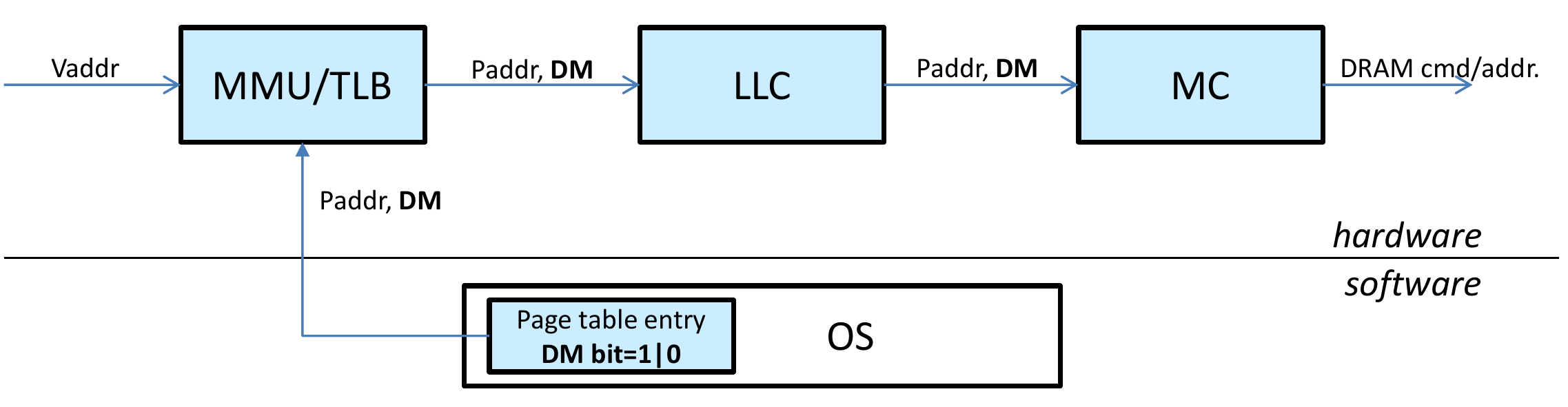}
  \caption{Deterministic memory-aware memory hierarchy: Overview}
  \label{fig:hw-process}
\end{figure}

Figure~\ref{fig:hw-process} shows this information flow of deterministic
memory. Note that bus protocols (e.g., AMBA~\cite{arm2011axi_ace})
also should provide a mean to pass the deterministic memory
information into each request packet. In fact, many existing
bus protocols already support some forms of priority information as
part of bus transaction messages~\footnote{For example, ARM AXI4
  protocol includes a 4-bit QoS identifier AxQOS signal~\cite{arm2011axi_ace}
  that supports up to 16 different 
  priority classes for bus transactions.}. These fields are currently
used to distinguish priority between bus masters (e.g., CPU vs. GPU
vs. DMA controllers). A bus transaction for deterministic memory can
be incorporated into these bus protocols, for example, as a special
priority class. 
The deterministic memory information can then be utilized in
mapping and scheduling decisions made by the respective hardware
components in the memory hierarchy.

In the following, we focus on cache and DRAM controllers and how the
deterministic memory information can be utilized in these important
shared hardware resources. 

\subsection{Deterministic Memory-Aware Shared Cache} \label{sec:dm-cache}

In this subsection, we present a deterministic memory-aware shared cache
design that provides the same isolation benefits of traditional way-based
cache partitioning techniques while achieving higher cache space utilization.


{\bf Way-based Cache Partitioning:}
In a standard way-based partitioning, which is supported in several 
COTS multicore processors~\cite{e500mc,armarm}, each core 
is given a subset of cache ways. When a cache miss
occurs, a new cache line (loaded from the memory) is allocated on one
of the assigned cache ways in order not to evict useful cache lines of
the other cores that share the same cache set.
An important shortcoming of way-partitioning is, however, that its
partitioning granularity is coarse (i.e., way granularity) and the
cache space of each partition may be wasted if it is underutilized. 
Furthermore, even if fine-grain partition adjustment is possible, it
is not easy to determine the ``optimal'' partition size of a task
because the task's behavior may change over time or depending on the
input. As a result, it is often a common practice to conservatively
allocate sufficient amount of resource (over-provisioning), which will
waste much of the reserved space most of the time.

{\bf Deterministic Memory-Aware Replacement Algorithm:}
We improve way-based partitioning by taking
advantage of the deterministic memory abstraction. 
The basic approach is that we use way partitioning only for
deterministic memory accesses while allowing best-effort memory
accesses to use all the cache ways that do not currently hold
deterministic cache lines. 

\begin{figure} [h]
  \centering
  \includegraphics[width=0.9\textwidth]{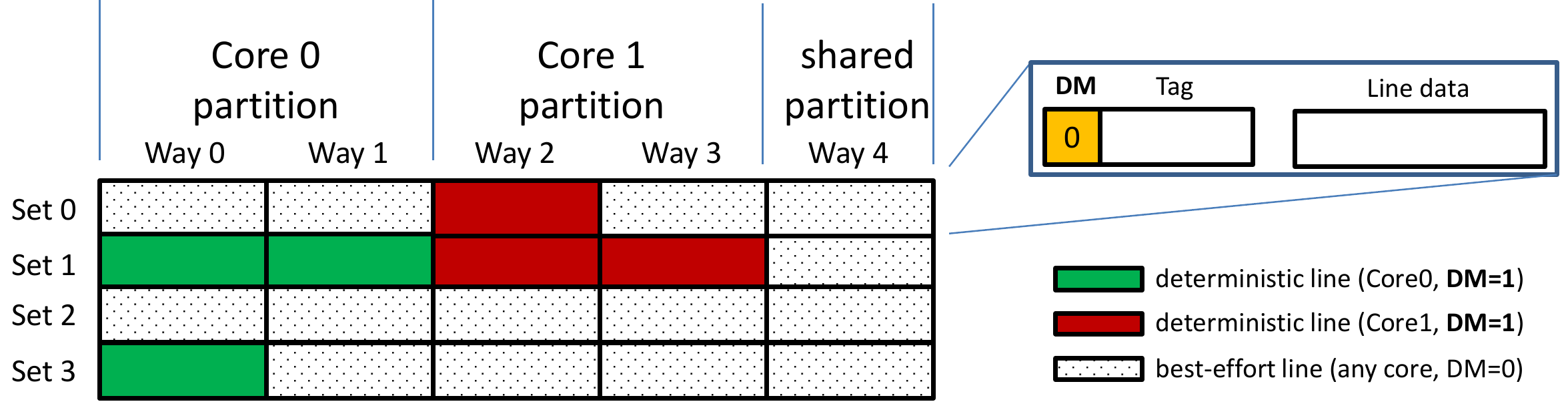}
  \caption{Deterministic memory-aware cache management}
  \label{fig:dm-cachepart}
\end{figure}

Figure~\ref{fig:dm-cachepart} shows an example cache status of our
design in which two cores share a 4-set, 5-way set-associative cache. 
In our design, each cache-line includes a $DM$ bit to indicate whether
the cache line is for deterministic memory or best-effort memory (see
the upper-right side of Figure~\ref{fig:dm-cachepart}). 
When inserting a new cache line (of a
given set), if the requesting memory access is for deterministic
memory, then  the victim line is chosen from the core's way partition
(e.g., way 0 and 1 for Core 0 in Figure~\ref{fig:dm-cachepart}).
On the other hand, if the requesting memory access is for best-effort
memory, the victim line is chosen from the ways that do not hold
deterministic cache lines. (e.g., in set 0 of
Fugre~\ref{fig:dm-cachepart}, all but way 2 are best-effort cache
lines; in set 1, 
only the way 4 is best-effort cache line.)

\begin{algorithm}[h]
  \SetKwInOut{Input}{Input}
  \SetKwInOut{Output}{Output}
  \Input{$PartMask_i$ - way partition mask of Core $i$ \newline $DetMask_s$ -
    deterministic ways of Set $s$}
  \Output{$victim$ - the victim way to be replaced. }

  \SetAlgoLined\DontPrintSemicolon
  \eIf{ $DM$ == 1 }{
    \eIf{ $(PartMask_i \land \neg DetMask_s) \neq NULL$ } { 
      \tcp{evict a best-effort line first}
      $victim = LRU(PartMask_i \land \neg DetMask_s)$\;
      $DetMask_s \mid = 1 \ll victim$\;
    }{
      \tcp{evict a deterministic line}
      $victim = LRU(PartMask_i)$
    }
  }{
    \tcp{evict a best-effort line}
    $victim = LRU(\neg DetMask)$ \;
  }
  \KwRet $victim$\;

  \caption{Deterministic memory-aware cache line replacement
    algorithm.}
  \label{alg:cacherepl}
\end{algorithm}

Algorithm~\ref{alg:cacherepl} shows the pseudo code of the cache line
replacement algorithm. As in the standard cache way-partitioning, we
assign dedicated cache ways for each core, denoted as $PartMask_i$, to
eliminate inter-core cache interference.  Note that $DetMask_s$
denotes the bitmask of the set $s$'s cache lines that contain
deterministic memory.
If a request from core $i$ is a deterministic memory request ($DM = 1$),
then a line is allocated from the core's cache way partition ($PartMask_i$). 
Among the ways of the
partition, the algorithm first tries to evict a best-effort
cache line, if such a line exists (Line 3-4). 
If not (i.e., all lines are deterministic ones), it chooses one of the
deterministic lines as the victim (Line 6).
One the other hand, if a best-effort memory is
requested ($DM \neq 1$), it evicts one of the best-effort cache lines,
but not any of the deterministic cache lines (Line 9). In this way,
while the deterministic cache lines of a partition are completely
isolated from any accesses other than the assigned core of the
partition, any under-utilized cache lines of the partition can still
be utilized as best-effort cache lines by all cores.




{\bf Deterministic Memory Cleanup:}\label{sec:dm_cleanup}
Note that a core's way partition would
eventually be filled with deterministic cache lines (ones with $DM =
1$) if left unmanaged (e.g., scheduling multiple different real-time tasks on the
core). This would eliminate the space efficiency gains of using
deterministic memory because the deterministic
memory cache lines cannot be evicted by best-effort memory requests.

In order to keep only a
minimal number of deterministic cache lines on any given partition in
a predictable manner, our cache controller provides a special
hardware mechanism that clears the $DM$ bits of all deterministic
cache lines, effectively turning them into best-effort cache-lines.
This mechanism is used by the core's OS scheduler on each
context switch so that the deterministic cache-lines of the previous
tasks can be evicted by the current task.
When the deterministic-turned-best-effort cache-lines of a task are
accessed again and they still exist in the cache, they will be
simply re-marked as deterministic without needing to reload from
memory. 
In the worst case, however, all deterministic cache lines of a task
shall be reloaded when the task is re-scheduled on the CPU.

Note that our cache controller reports the number of deterministic
cache lines that are cleared on a context switch back to the OS. This
information can be used to more accurately estimate cache-related
preemption delays (CRPD)~\cite{altmeyer2012improved}.


{\bf Guarantees:}
The premise of the proposed cache replacement strategy is
that a core's deterministic cache lines will never be evicted by other
cores' cache allocations, hence \emph{preserving the benefit of cache
partitioning}. At the same time, non-deterministic 
cache lines in the core's cache partition can safely be used as other
cores' best-effort memory requests, hence \emph{minimizing wasted
 cache capacity} due to partitioning.

{\bf Comparison with PRETI:}
Our DM-aware cache replacement algorithm is similar to several prior
mixed-criticality aware cache
designs~\cite{lesage2012preti,kumar2014cache,yan2011time}.
The most closely related work is
PRETI~\cite{lesage2012preti}, which also modifies LRU to be
mixed-criticality-aware.
There are, however, several notable differences.
First, PRETI uses the thread(task)-id to distinguish critical and
non-critical cache accesses, whereas we use MMU, which enables finer,
page granularity criticality control. 
Second, in PRETI, cache-lines reserved for a critical thread can
only be released on its termination. In contrast, we provide a
DM-cleanup mechanism, which enables efficient reclamation of
deterministic memory cache-lines at each context-switch.
Third, PRETI's replacement algorithm provides a firm cache space
reservation capability~\cite{Raj98} in the sense that a real-time task
can utilize more than its dedicated
private space, whereas our replacement algorithm does not allow such
additional cache space utilization.
Lastly, while the prior works mainly focus on the cache, our main
goal is to provide a unified framework---the deterministic memory
abstraction---which can carry information about time-sensitivity of
memory space not only 
in the cache, but also in the OS and throughout the entire memory hierarchy.
We will discuss how a traditional DRAM controller can be extended to
support deterministic memory abstraction in the following.

\subsection{Deterministic Memory-Aware DRAM Controller}\label{sec:dm-dram}


In this subsection, we present a deterministic memory-aware DRAM
controller design, which, in collaboration with our OS support,
provides strong spatial and temporal isolation for 
deterministic memory accesses while also enables efficient best-effort
memory processing. 

\begin{figure}[h]
\centering
  \begin{subfigure}{0.45\textwidth}
    \includegraphics[width=\linewidth,height=0.2\textheight]{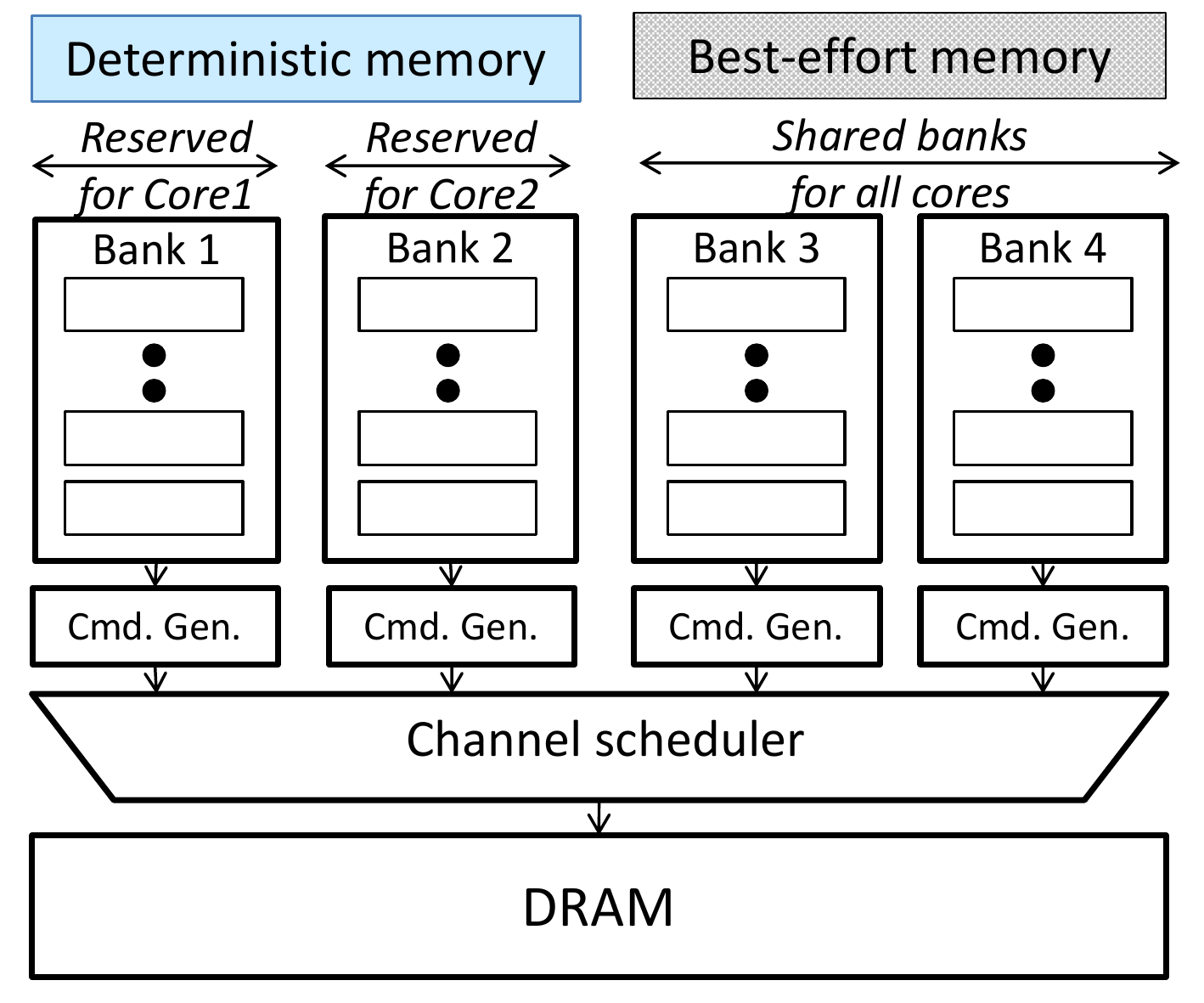}
    \caption{MC architecture}
    \label{fig:medusa-arch}
  \end{subfigure}
  \begin{subfigure}{0.45\textwidth}
    \includegraphics[width=\linewidth,height=0.2\textheight]{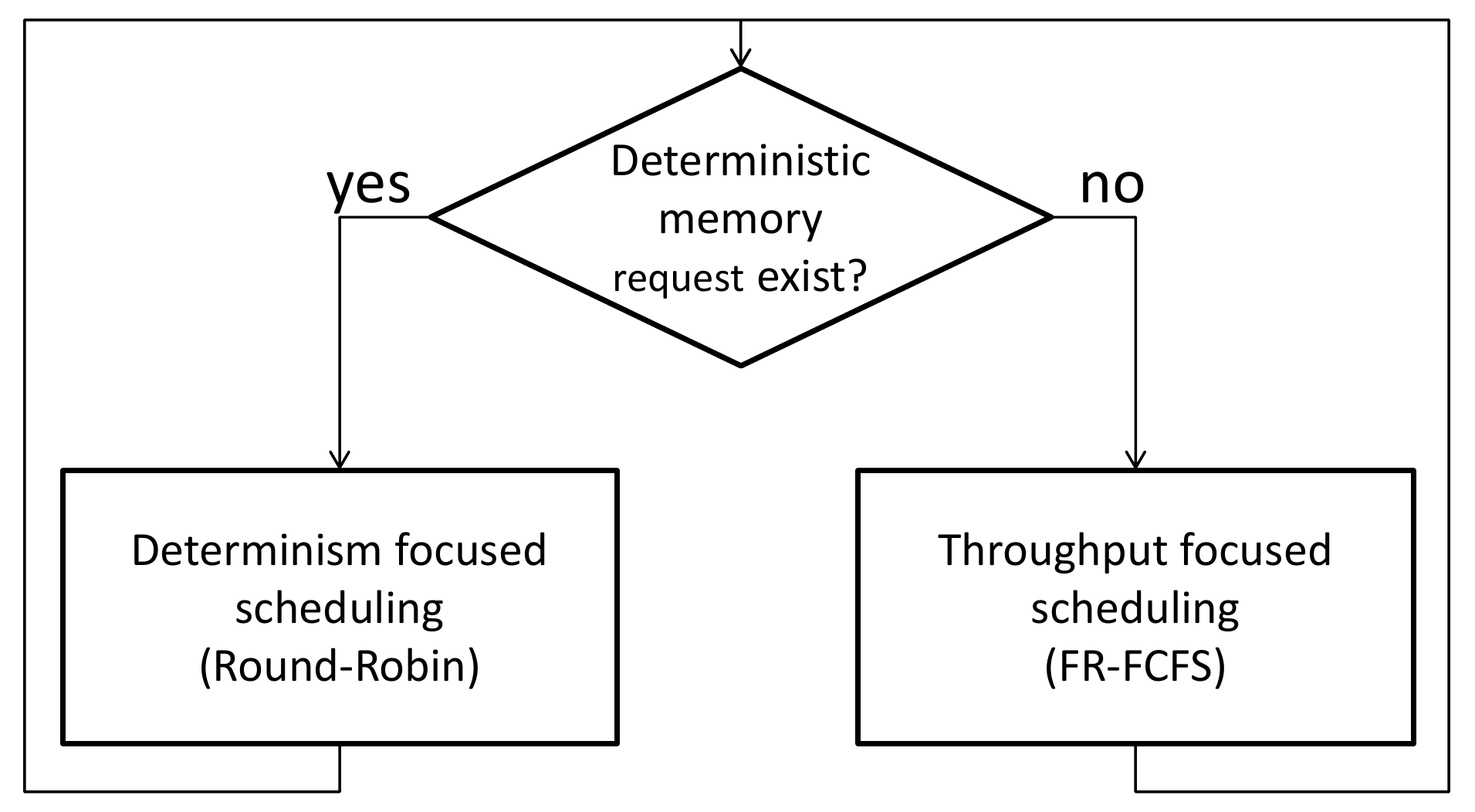}
    \caption{Scheduling algorithm}
    \label{fig:medusa-schedule}
  \end{subfigure}
  \caption{Deterministic memory-aware memory controller architecture
    and scheduling algorithm. }  \label{fig:timing-reorder}
\end{figure}

First, the OS actively controls on which DRAM bank a page frame
is allocated.
Specifically, the OS reserves a small number of banks for each core to
be used as the deterministic memory for the core, while the rest of the
banks are used for the best-effort memory of all cores, as shown in
Figure~\ref{fig:medusa-arch} (also in Figure~\ref{fig:osview}).
When the OS allocates memory pages of an
application task, deterministic memory pages of the task shall be
allocated on core-private DRAM banks to eliminate DRAM bank-level
inter-core interference~\cite{kim2014rtas,valsan2015cpsna}, while
best-effort memory pages are allocated on the shared DRAM banks.

Second, the memory controller (MC) implements a \emph{two-level
 scheduling} algorithm that first prioritizes deterministic memory
requests over the ones for best-effort memory.
For deterministic memory requests, we use a round-robin
scheduling policy as it offers stronger time
predictability~\cite{paolieri2009analyzable}, while we use
first-ready first-come-first-serve (FR-FCFS) policy for
scheduling best-effort memory requests as it offers
high average throughput~\cite{rixner2000memory}.
Figure~\ref{fig:medusa-schedule} shows the flowchart of the scheduler.
Note, however, that strictly prioritizing deterministic memory requests
could starve best-effort memory requests indefinitely. Since we assume the
existence of a pessimistic worst-case bound for best-effort memory, we
limit the maximum number of consecutive processing of deterministic 
memory requests in case best-effort memory requests exist, in order to
achieve tightly bounded worst-case timing for
deterministic memory while achieving pessimistic, but still bounded,
worst-case timing for best-effort memory.

Our design is inspired by prior mixed-criticality memory controller
proposals~\cite{kim2015predictable,jalle2014dual,valsan2015cpsna}, all
of which, like us, apply different scheduling algorithms
depending on memory criticality, although detailed designs (and
assumptions) are varied. In this work, we particularly use the MEDUSA
memory controller design~\cite{valsan2015cpsna} as our baseline, but 
improve its efficiency by leveraging the DM-bit information passed
down to the memory controller. Specifically,
in~\cite{valsan2015cpsna}, a real-time task has to allocate its entire 
memory space from the reserved DRAM banks, even when much of its
allocated memory is never used in the time-critical part. In contrast,
our design can \emph{reduce} the amount of memory allocated in the reserved
DRAM banks by only allocating the deterministic memory pages. This
allows us to accommodate more real-time tasks with the same amount of
reserved DRAM banks.

The necessary changes to support deterministic memory is
small. Specifically, the original MEDUSA
controller~\cite{valsan2015cpsna} uses a set of memory 
controller specific hardware registers to identify reserved DRAM
banks of the cores. Instead, our modified memory 
controller design simply uses the DM-bit information in each memory
request to determine memory criticality. Other mixed-criticality
real-time  memory controllers
designs~\cite{kim2015predictable,jalle2014dual} also similarly rely on
memory-controller-specific hardware registers to identify memory
criticality. Thus, we believe they also can be easily augmented to
support the deterministic memory abstraction.

\subsection{Other Shared Hardware Resources}

We briefly discuss other potential deterministic memory-aware shared
hardware designs.

As shown in ~\cite{valsan2016taming}, the 
miss-status-holding-registers (MSRHs) in a shared non-blocking cache
can be a significant source of inter-core interference if the number
of MSHRs in the shared cache is insufficient to support the memory
parallelism of the cores. The contention in MSHRs can be avoided by simply
having a sufficient number of MSHRs, as we did in our evaluation setup.
But if it is
difficult for the reasons discussed in~\cite{valsan2016taming}, 
deterministic memory-aware MSHR management can be alternatively
considered. For example, one possible DM-aware approach is that
reserving some per-core MSHR entries to handle deterministic memory
and sharing the rest of MSHR entries for best-effort memory requests
from all cores.

Deterministic memory-aware TLB can also be considered. Although a TLB
is not typically shared among the cores, it is conceivable to design a
DM-aware TLB replacement policy that reserves some TLB entries for
deterministic memory that cannot be evicted by access to best-effort
memory addresses. Such a policy can be useful to reduce task WCET and
CRPD overhead within a core.

\section{Timing Analysis}~\label{sec:analysis}

In this section, we show how the traditional response time
analysis (RTA)~\cite{Audsley93RTA} can be extended to account for
deterministic and best-effort memory abstractions.

In our system, a real-time task $\tau_i$ is represented by the
following parameters:
\begin{equation}
  \tau_i = \{C_i, T_i, D_i, DM_i, BM_i\}
\end{equation}

where $C_i$ is the WCET of the $\tau_i$ when it executes in
isolation; $T_i$ is the period of the task; $D_i$ is the deadline of
the task; 
$DM_i$ represents the maximum number of deterministic memory
requests that suffer inter-core interference; $BM_i$ is the
maximum number of best-effort memory requests that are subject to
inter-core interference.

Note that all these parameters can be obtained in \emph{isolation}.
A task is said to execute
in isolation if: (1) it executes alone on the assigned core under a
given resource partition; and (2) all the other $N_{proc} - 1$ cores
are idle or offline.

Note also that in the proposed DM-aware system described earlier, $DM_i$
accounts only a subset of deterministic memory accesses that
result in L2 misses because the L2 hit accesses would not suffer
inter-core interference. On the other hand, $BM_i$ would represent a
subset of best-effort memory accesses that result in L1 misses (not L2
misses). This is because for best-effort memory, L2 cache space is
shared, and, in the worst-case, all of them will have to be fetched
from the memory controller.




We then can compute $\tau_i$'s worst-case memory interference delay
$I_i$ as follows:
\begin{equation}
  I_i = DM_i \times RD^{dm} + BM_i \times RD^{bm},
\label{eq:interference}
\end{equation}

where $RD^{dm}$ and $RD^{bm}$ denote the \emph{worst-case inter-core
  interference delay} of a deterministic and best-effort memory
request, respectively.

By our system design, $RD^{dm}$ is small and tightly bounded because
accesses to deterministic memory suffer minimal (or zero) inter-core
interference at the shared L2 cache and the shared DRAM.
On the other hand, $RD^{bm}$ will be substantially higher and highly
pessimistic because we have to pessimistically assume access to
best-effort memory will always miss the L2 cache and suffer high
queuing delay at the DRAM controller. 

Traditional RTA analysis can then be performed by finding the first
value of $k$ such that $R^{(k+1)}_i = R^{(k)}_i$ (task is schedulable)
or such that $R^{(k)}_i > D_i$ (task is not schedulable), given that
$R^{(0)}_i = C_i + I_i$ and that $R^{(k+1)}_i$ is calculated as:


\begin{equation}
  R_i^{(k+1)}\!\!~=~\!\! C_i + I_i + \!\! \sum_{\tau_j \in hp(i)}
  \left\lceil\frac{R_i^{(k)}}{T_j}\right\rceil \cdot (C_j + I_j),
\end{equation}

where $hp(i)$ is the set of all the tasks with priority higher than
$\tau_i$.

The major benefit of our approach is its flexibility.
For example, a pure COTS multicore system may provide high performance
but, doesn't provide isolation guarantees. Therefore, all access to
shared resource may need to be assumed to suffer highly pessimistic
worst-case inter-core interference delay (e.g., $RD^{bm}$ above)
because no isolation is guaranteed.  
On the other hand, a fully time-predictable hardware
architecture~\cite{liu2012pret,merasa2010micro,ungerer2013parmerasa}
may provide strong timing predictability with a small tight worst-case
inter-core interference delay (e.g., $RD^{dm}$ above), but not high
performance and efficiency.
In contrast, the flexibility of our approach enables hardware
designs that optimize differently depending on the memory type, which
in turn enables analyzable and efficient multicore systems.

\section{Prototype Implementation} \label{sec:impl}

In this section, we provide
implementation details of our prototype, which is based on Linux
3.13 kernel and gem5~\cite{binkert2011gem5} full-system simulator.
First, we briefly review the
ARMv7 architecture on which our implementation is based
(Section~\ref{sec:armbg}). We then 
describe our modifications to the Linux kernel to support the
deterministic memory abstraction
(Section~\ref{sec:impl-linux}). Lastly, we describe the hardware 
extensions on the gem5 simulator (Section~\ref{sec:impl-hw}).

\subsection{ARM Architecture Background}\label{sec:armbg}
\begin{figure}[h]
  \centering
  \includegraphics[width=\linewidth]{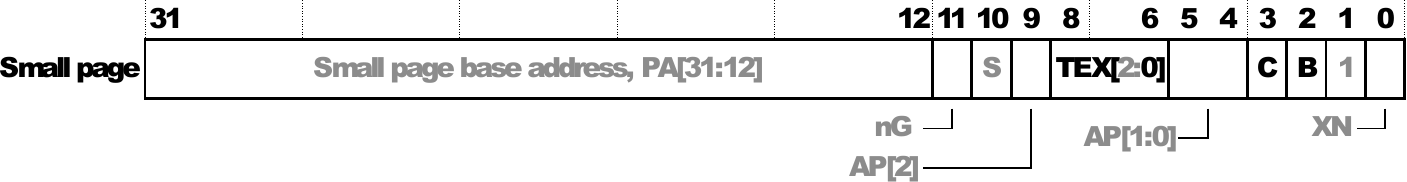}
  \caption{Small descriptor format for 2$^{nd}$ level page table entry
    in ARMv7-A family SoCs~\cite{armarm}.}
  \label{fig:pgtable_fmt}
\end{figure}

We use the ARMv7-A~\cite{armarm} architecture because it is well
supported by the gem5 simulator.
The ARMv7 architecture defines four primary memory types and
several memory-related attributes such as cache policy
(write-back/write-through) and coherence boundaries (between cores or
beyond). Up to 8 different combinations are allowed by the
architecture. Each page's memory type is determined by a set of bits
in the corresponding 2$^{nd}$-level page table
entry. Figure~\ref{fig:pgtable_fmt} illustrates the structure of a
page table entry (PTE).


In the figure, the bits \texttt{TEX[0]}, \texttt{C} and \texttt{B} are
used to define one of the 8 memory types. The property of each memory
type is determined by two global architectural registers, namely
Primary Region Remap Register (PRRR) and Normal Memory Region Register
(NMRR)~\footnote{The hardware behaves as described only when the so
-called ``TEX remapping'' mechanism is in use. TEX remapping can be
  controlled via a configuration bit (TRE) in the System Control
  Register (SCTLR). The Linux kernel enabled TEX remapping by
  default.}. 


\subsection{Linux Extensions} \label{sec:impl-linux}


We have modified Linux kernel 3.13 to support deterministic
memory. 

At the lowest level, we define a new memory
type that corresponds to the deterministic memory. 
The default ARM Linux uses only 6 out of 8 possible memory types of
ARMv7, leaving two undefined memory types. For deterministic memory, we
define one of the unused memory types as the deterministic memory
type, by updating PRRR and NMRR registers at boot time. 
A page is marked as deterministic memory when the corresponding page
table entry's memory attributes point to the deterministic memory
type.

At the user-level, we extend Linux's ELF (Executable and
Linkable Format~\cite{elf1995tool}) loader and the {\tt exec} system call
implementation. We currently use a special file extension to inform
the ELF loader whether to mark the entire memory address or a subset
of task's memory pages as deterministic memory.
For fine-grained control, the virtual page numbers which might be
marked as deterministic memory are currently hard-coded in the kernel
source and a subset of them is selected based on the arguments
passed to the {\tt exec} system call. In the future, we will use Linux
kernel's \texttt{debugfs} interface to efficiently communicate page information.
Also, the ELF header of a program binary can be used instead to encode
the virtual page numbers.

Within the Linux kernel, a task's virtual address space is represented
as a set of \emph{memory regions}, each of which is represented by a
data structure, \texttt{vm\_area\_struct}, called a VMA
descriptor. Each VMA descriptor contains a variety of metadata about
the memory region, including its memory type information. Whenever a
new physical memory block is allocated (at a page fault), the kernel
uses the information stored in the corresponding VMA descriptor to
construct the page table entry for the new page. We add a new flag
\texttt{VM\_DETMEM} to indicate the deterministic memory type in a VMA
descriptor. When a page fault happens on accessing a memory address, if the
\texttt{VM\_DETMEM} flag of the memory region corresponding to the address
is set, or the address falls within one of the virtual page numbers
hard-coded in the kernel
then OS sets the \texttt{TEX[0]}, \texttt{C} and \texttt{B} bits in
allocating the page for the address to mark that it is a
deterministic memory page.

Note that the above code changes are minimal. In total, we
only have added/modified less than 200 lines of C and assembly code
over 12 files in the Linux kernel source tree. Furthermore, because
most changes are in page table descriptors and their initialization,
no runtime overhead is incurred by the code changes.

We then have applied the PALLOC patch~\cite{yun2014rtas}, which replaces
the buddy allocator to support DRAM bank-aware page allocation. We
further extend the PALLOC allocator to support deterministic
memory. Specifically, we extend PALLOC's \texttt{cgroup} interface to declare
a subset of banks to be used as private banks for the cgroup's
deterministic memory pages and another subset of banks to be used for
best-effort memory pages.

\subsection{Gem5 Extensions}\label{sec:impl-hw}

We have modified the gem5 full-system simulator as follows.

{\bf MMU and TLB.}
The deterministic memory type information stored in the page table is
read by the MMU and passed throughout the memory hierarchy. 
When a page fault occurs, the MMU performs the page table walk to
determine the physical address of the faulted virtual address. In the
process, it also reads other important auxiliary information such as
memory attribute and access permission from the page table entry and
stores them into a TLB entry in the processor. The deterministic memory
attribute is stored alongside with the other memory attributes in the TLB
entry.  
Specifically, we add a single bit in the gem5's implementation of a
TLB entry to indicate the deterministic memory type.
As a reference, Cortex-A17's TLB entry has 80 bits and
a significant fraction of the bits are already used to store various
auxiliary information~\cite{arm-a17} or reserved for future use. Thus,
requiring a single bit in a TLB entry does not pose significant
overhead in practice. We also extend the memory
request packet format in the gem5 simulator to include the
deterministic memory type information. In this way, the memory type
information of each memory request can be passed down through the
memory hierarchy. 
In real hardware, bus protocols should be extended to include such
information. As discussed earlier, existing bus protocols such as AXI4
already support the inclusion of such additional information in each bus
packet~\cite{arm2011axi_ace}.


{\bf Cache Controller.}
The gem5's cache subsystem implements a flexibly configurable
non-blocking cache architecture and supports standard LRU and random
replacement algorithms.
Our modifications are as follows. First, we extend gem5's cache
controller to support a standard way-based partitioning capability~\footnote{https://github.com/farzadfch/gem5-cache-partitioning}. 
The way partition is configured via a set of programmable registers.
When a cache miss occurs, instead  
of replacing the cache line in the LRU position, the controller
replaces the LRU line among the configured ways for the core. The
way-based partitioning mechanism is used as a baseline. 
On top of the way-based partitioning, we implement the proposed
deterministic memory-aware replacement and cleanup 
algorithms (Section~\ref{sec:dm-cache}).

{\bf DRAM Controller.}
Gem5's memory controller subsystem supports a standard
FR-FCFS algorithm~\cite{hansson2014simulating}. We have extended the
memory controller subsystem to support the two-level scheduling
algorithm described in~\cite{valsan2015cpsna}. The two-level scheduler
is modified to leverage the DM bit passed to the memory
controller as part of each memory request bus transaction. Also, to
prevent starvation of best-effort memory requests, we limit the
maximum consecutive deterministic memory request processing to 30
when one or more best-effort memory requests are in the
memory controller's queue.

\section{Evaluation} \label{sec:evaluation}

In this section, we present evaluation results to support the
feasibility and effectiveness of the proposed deterministic
memory-aware system design.


{\bf System Setup.} For OS, we use a modified Linux kernel 3.13, which
implements the 
modifications explained in Section~\ref{sec:impl-linux} to support the
deterministic memory abstraction. For hardware, we use a
modified gem5 full system simulator, which implements the proposed
deterministic memory support described in Section~\ref{sec:impl-hw}.
The simulator is configured as a quad-core out-of-order
processor (O3CPU model~\cite{o3cpu}) with per-core private L1 I/D
caches, a shared L2 cache, and a shared DRAM. The baseline architecture
parameters are shown in Table~\ref{tbl:sysconf}.
We use the \texttt{mlockall} system call to allocate all necessary
pages of each real-time application at the beginning so as to avoid
page faults during the rest of program's execution. In addition, we
enabled the kernel configuration option \texttt{NO\_HZ\_FULL} to
reduce unnecessary scheduler-tick interrupts.



\begin{table} [h]
  \centering
  \caption{Simulator configuration}
  \begin{tabular}{|c|c|}
    \hline
    Core            & Quad-core, out-of-order, 2 GHz, IQ: 96, ROB: 128, LSQ: 48/48	\\
    \hline
    L1-I/D caches    & Private 16/16 KiB (2-way), MSHRs: 2(I)/6(D) \\
    \hline
    L2 cache        & Shared 2 MiB (16-way), LRU, MSHRs: 56, hit latency: 12 \\
    \hline
    DRAM Controller & Read buffer: 64, write buffer: 64, open-adaptive page policy \\  
    \hline
    DRAM module     & LPDDR2@533MHz, 1 rank, 8 banks\\
    \hline
  \end{tabular}
  \label{tbl:sysconf}
\end{table}


\subsection{Real-Time Benchmark Characteristics} \label{sec:profile}


We use a set of EEMBC~\cite{eembc} automotive and
SD-VBS~\cite{venkata2009sd} vision benchmarks (input: sim) as
real-time workloads.
We profile each benchmark, using the gem5
simulator, to better understand memory
characteristics of the benchmarks.



\begin{figure} [h]
    \centering
    \begin{subfigure}{0.48\textwidth}
        \includegraphics[width=\linewidth]{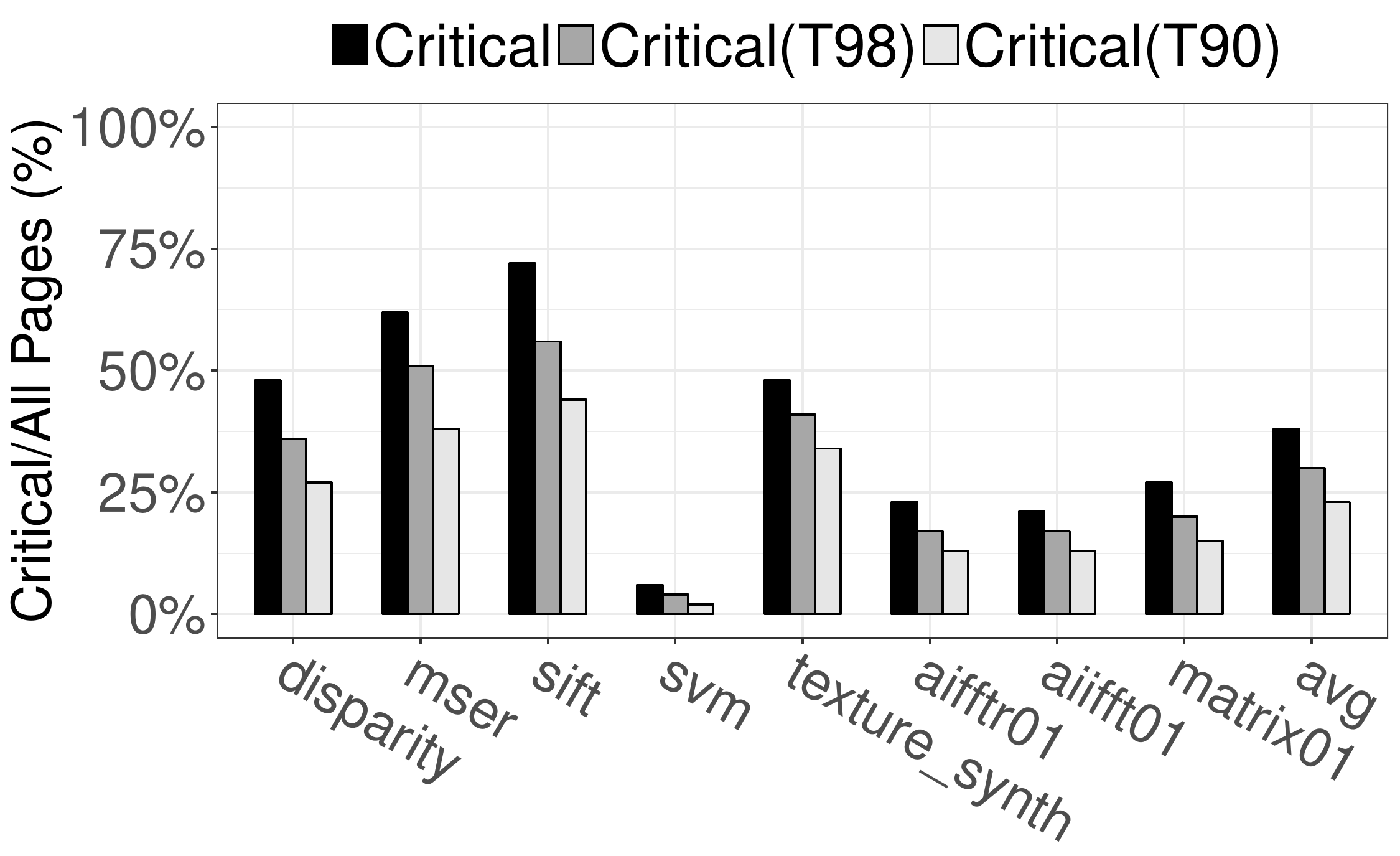}
        \caption{Critical pages among all touched pages.}
        \label{fig:n-pages}
    \end{subfigure}
    \hfill
    \begin{subfigure}{0.48\textwidth}
        \includegraphics[width=\linewidth]{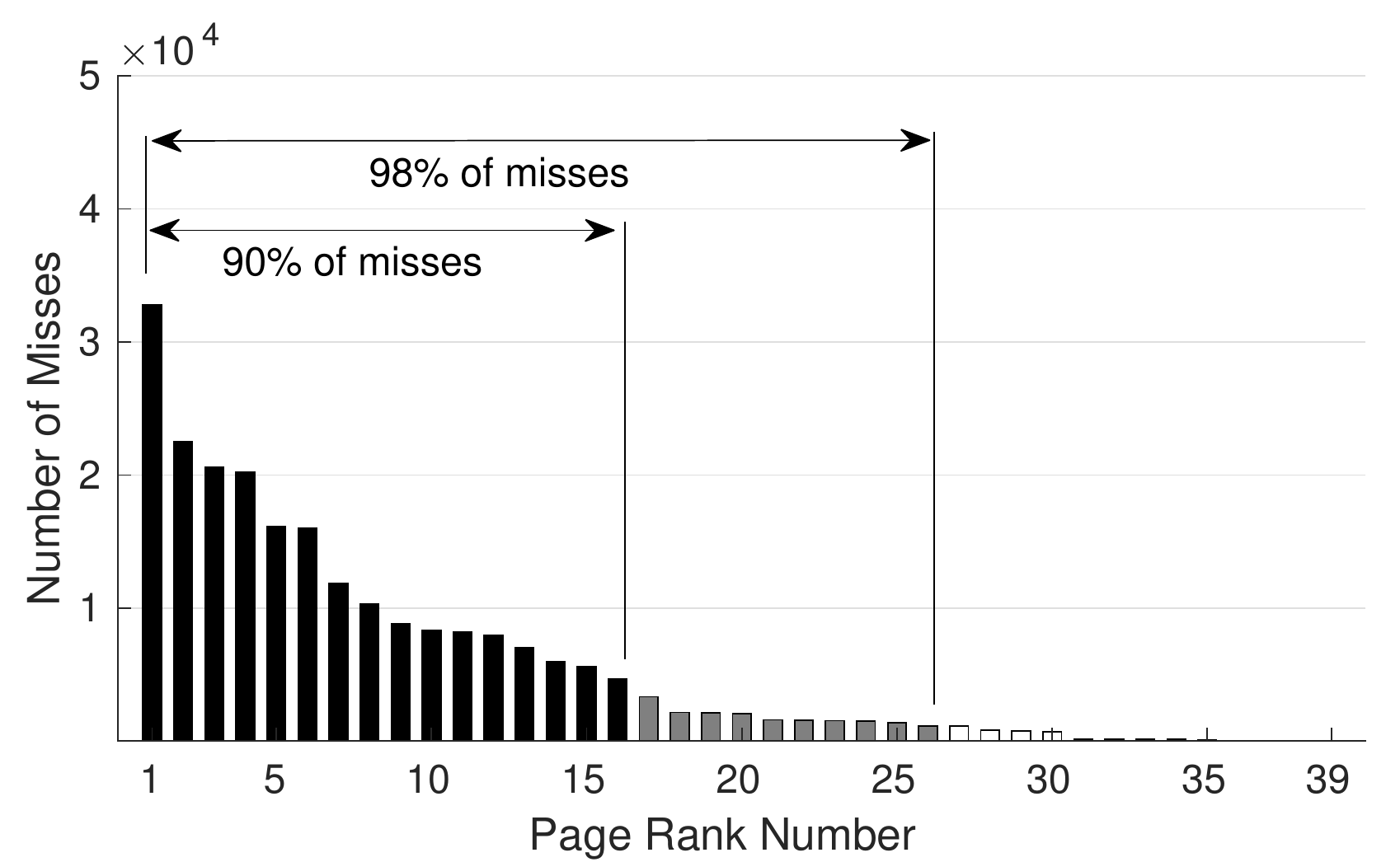}
        \caption{\emph{svm} L1 miss distribution of critical pages.}
        \label{fig:svm-l1missdist}
    \end{subfigure}

    \caption{Space and temporal characteristics of application memory
    pages. Critical pages refer to the touched pages within the main loop 
    of each benchmark.} 
\end{figure}

Figure~\ref{fig:n-pages} shows the ratio between the number of
accessed pages within the main loop and the number of all accessed
pages of each benchmark; the pages accessed in the loop are denoted
as \emph{critical pages}. To further analyze the characteristics of the
critical pages, we profiled L1 cache misses of each critical page to
see which pages contribute most to the overall L1 cache
misses. \emph{Critical(T98)} shows the ratio of ``top'' critical
pages which contribute to 98\% of the L1 cache misses.
The same is for \emph{Critical(T90)} except that 90\% of the L1 cache
misses are considered.
As can be seen in the figure, only 38\% of all pages, on average, are
critical pages, and this number can be as low as 6\% (\emph{svm}.)
This means that the rest of the pages are accessed during the
initialization and other non-time-critical procedures. 
This ratio is further reduced to 23\% of the touched pages if 90\% of
L1 cache misses are considered. 

Note that in our system setup, the private L1 cache misses are
directed to the shared L2 cache, which is   
shared by all cores. Thus, those pages that show high L1 misses 
likely contribute most to the WCET of the application because they
can suffer from high inter-core interference due to contention at the
shared L2 cache and/or the shared DRAM.
Figure~\ref{fig:svm-l1missdist} shows how we determine top critical
pages for the \emph{svm} benchmark. We rank all pages based on
the number of L1 cache misses of each page.
In case of \emph{svm}, the top 26 and 16 pages account for 98\% and
90\% of misses of all critical pages. These pages
are 4\% and 2\% of all the touched pages, respectively, as shown
in Figure~\ref{fig:n-pages}. This suggests even among the critical pages,
certain pages contribute more to WCET than the rest of the critical pages.

The results show that selective, fine-grained application of
deterministic memory can significantly reduce WCETs while minimizing
resource waste.

\subsection{Effects of Deterministic Memory-Aware Cache}\label{sec:result}

In this experiment, we study the effectiveness of the proposed deterministic
memory-aware cache. The basic experimental setup is that we run a
real-time task on Core~3 and 
three instances of a memory intensive synthetic benchmark
(\emph{Bandwidth} with write memory access pattern from the IsolBench
suite~\cite{valsan2016taming}) as best-effort 
co-runners on Core~0 through~2.
Note that the working-set size of the best-effort co-runners
is chosen so that the sum of all co-runners is equal to the size of
the entire L2 cache. This will increase the likelihood to evict the
cache lines of the real-time task if its cache lines are not
protected.

We evaluate the system with 5 different configurations: \emph{NoP, WP, 
DM(A), DM(T98) and DM(T90)}. In NoP, the L2 cache is shared among all
cores without any restrictions. In WP, the L2 cache is partitioned
using the standard way-based partitioning method, where 4 dedicated
cache ways are given to each core.
In DM(A), the entire address space of the real-time task is
marked as deterministic memory, while in DM(T98) and DM(90), only the
pages which account for 98\% and 90\%of the L1 misses, respectively,
of the task's critical pages are marked as deterministic.
In all DM configurations, each core is given 1/4 of the cache ways for
the core's deterministic memory.

Note that, in this experiment, the results for DM(A) will be similar
to that of PRETI~\cite{lesage2012preti}, because, in both systems, a
dedicated cache space is guaranteed to a real-time task's entire
memory space, while the presence of memory-intensive best-effort
co-runners would prevent the real-time task under PRETI from utilizing
additional cache space.



\begin{figure} [h]
    \centering
    \begin{subfigure}{0.48\textwidth}
        \includegraphics[width=\linewidth]{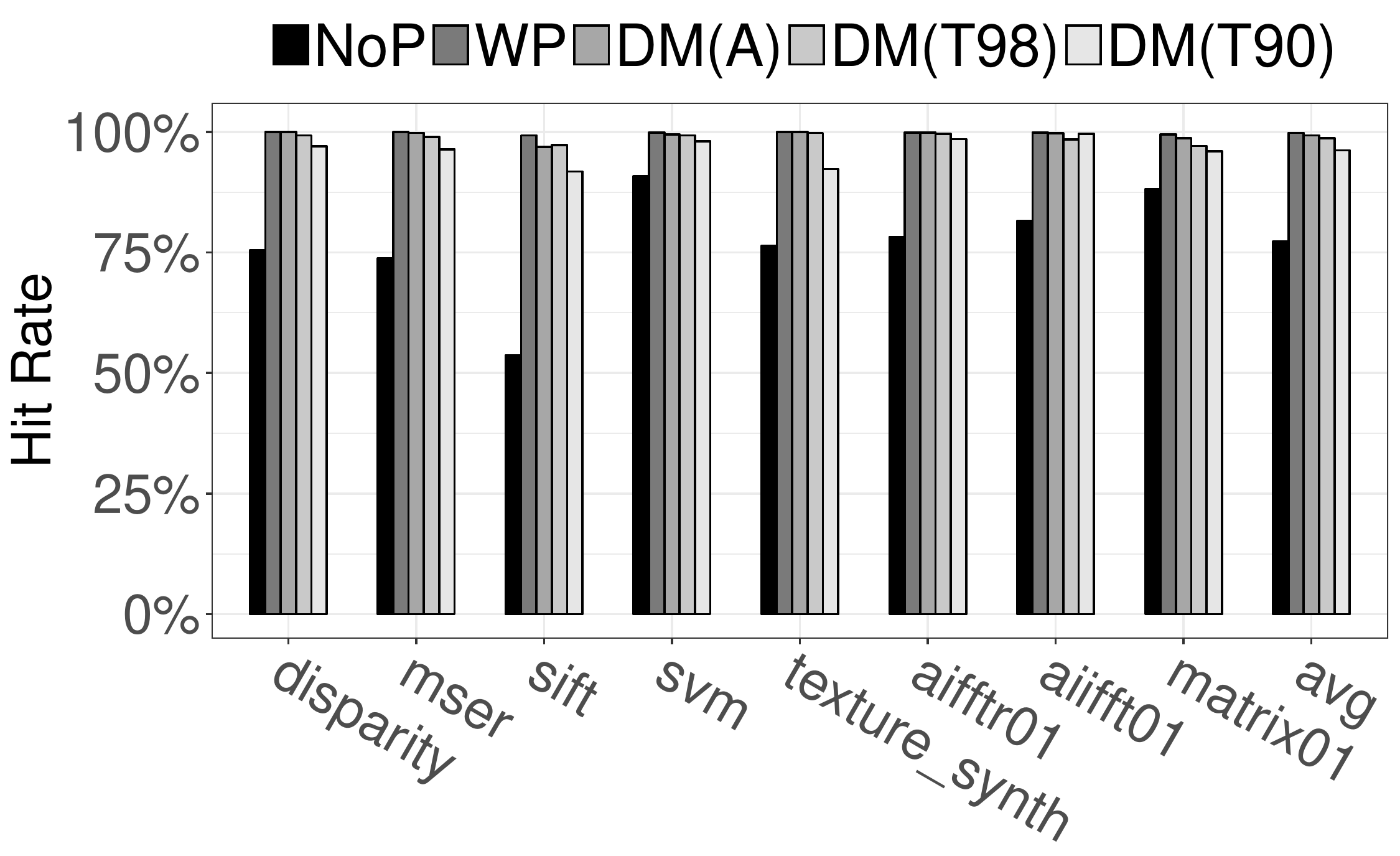}
        \caption{L2 hit rate.}
        \label{fig:hitrate}
    \end{subfigure}
    \hfill
    \begin{subfigure}{0.48\textwidth}
      \includegraphics[width=\linewidth]{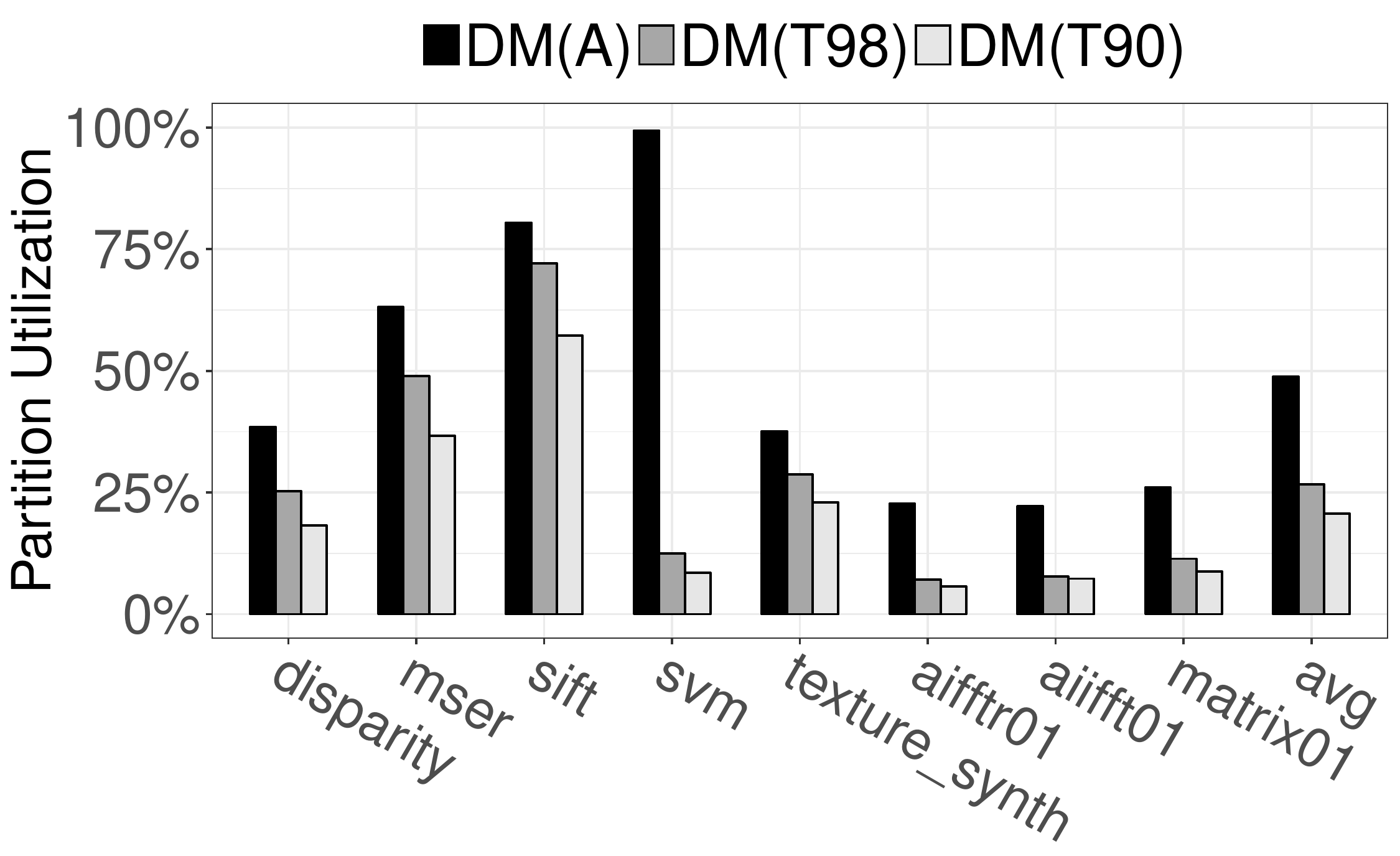}
      \caption{Cache partition usage of DM approaches.}
      \label{fig:dm-util}
    \end{subfigure}
    
    \caption{L2 hit-rate and cache space usage (deterministic memory
      only) of real-time tasks.} 
\end{figure}

{\bf Effects on Real-Time Tasks.} \label{subsec:real-time}
Figure~\ref{fig:hitrate} compares the L2 hit rates of real-time
tasks for each system configuration.
First, in NoP, the L2 hit rates are low (e.g., 54\% for \emph{sift})
because the cache lines of the real-time benchmarks are evicted by the
co-running \emph{Bandwidth} benchmarks.
In WP, on the other hand, all benchmarks show close to 100\% hit
rates. This is because the dedicated private L2 cache space (4 out of
16 cache ways = 512KB) is 
sufficient to hold the working-sets of the real-time benchmarks, which
cannot be evicted by the co-runners.
The hit rates are also close to 100\% in DM(A) because the co-runners
are not allowed to evict any of the cache lines allocated  
for the real-time tasks as their entire memory spaces (thus their
cache-lines in the L2) are marked as
deterministic memory. In DM(T98) and DM(90), not all pages are marked as
deterministic memory. As the result, the co-runners can evict some of
the best-effort cache lines of the real-time tasks and this in turn
results in slight reduction in the hit rates. 

Next, for all DM configurations, we measure the fraction of
deterministic memory cache-lines in a real-time task's cache partition
by checking $DM$ bit in the cache lines in the instrumented gem5
simulator.
Figure~\ref{fig:dm-util} shows the percentage of the cache
lines allocated by the deterministic memory cache lines.
On average, only 49\%, 27\%, and 21\% of cache-lines are deterministic
memory cache-lines for DM(A), DM(T98), and DM(T90), respectively. 
Note that when the conventional way
partitioning is used (in WP), the unused cache space in the private
cache partition is essentially wasted as no other task can utilize it.
In the deterministic memory-aware cache, on the other hand, the
best-effort tasks can use the non-DM cache lines in the cache partition.
Thus, the hit rate of the best-effort tasks can be
improved as more cache space will be available to them. This effect
will be shown in the following experiment. 

{\bf Effects on Best-Effort Tasks.} \label{subsec:best-effort}
To study the effect of deterministic memory-aware cache on realistic
best-effort tasks (as oppose the synthetic ones used above), we
designed an experiment with the \emph{bzip2} benchmark from SPEC2006
as the best-effort task running on Core 0, and 3 instances of a real-time
task running on Core 1 through 3. We chose \emph{bzip2} based on the
following selection criteria: 1) It must frequently access the
shared cache; 2) It must be sensitive to extra cache space (i.e. the
hit rate shall be improved if more cache space is given to the
benchmark). The \emph{bzip2} meet both requirements according to
a memory characterization study~\cite{jaleel2010memory} by Intel,
which is also confirmed in our simulation setup.


\begin{figure} [h]
    \centering
    \begin{subfigure}{0.48\textwidth}
        \includegraphics[width=\linewidth]{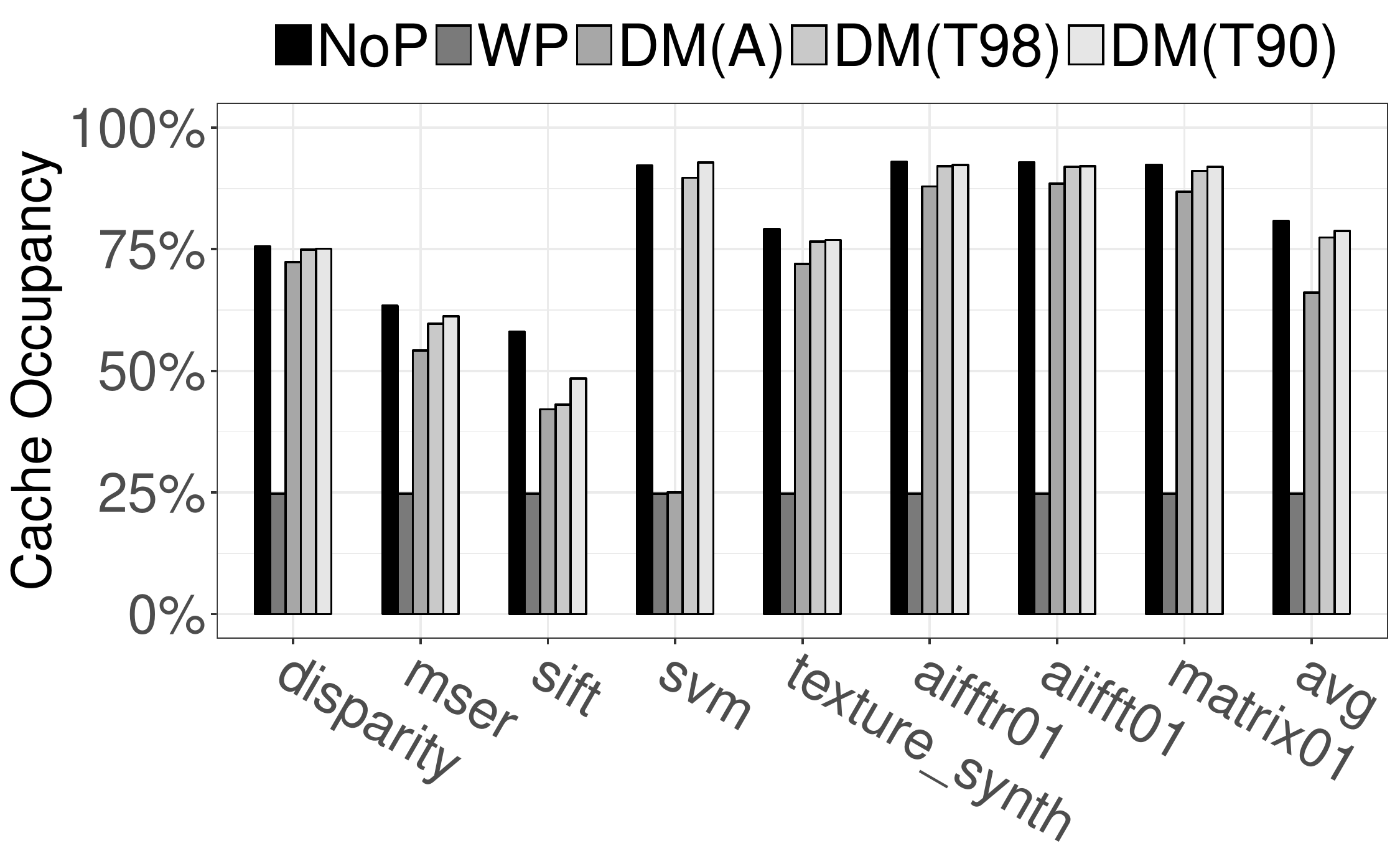}
        \caption{Cache space occupied by \emph{bzip2}.}
        \label{fig:bzip2-benefit_a}
    \end{subfigure}
    \hfill
    \begin{subfigure}{0.48\textwidth}
        \includegraphics[width=\linewidth]{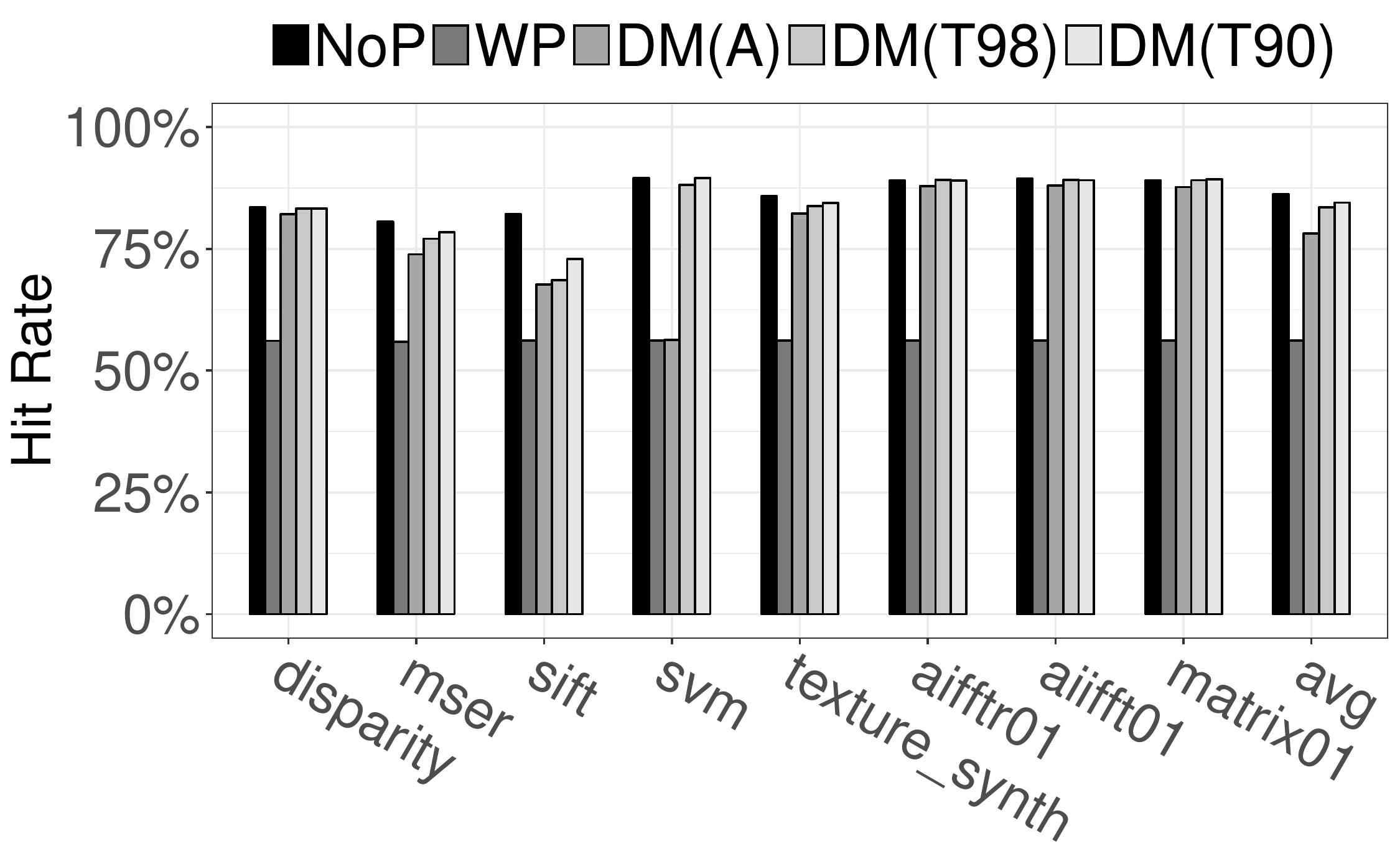}
        \caption{\emph{bzip2} hit rate.}
        \label{fig:bzip2-benefit_b}
    \end{subfigure}

    \caption{Cache usage and hit rate impact of DM-aware cache to the best-effort task
      (\emph{bzip2}).
    }
    \label{fig:bzip2-benefit}
\end{figure}

Figure \ref{fig:bzip2-benefit} shows the results.
Inset (a) shows the percentage of cache space used
by \emph{bzip2} for each real-time task pairing, while inset (b)
shows its hit rates.
Note that in WP, \emph{bzip2} can only use 25\% of cache space
(512kB out of 2MB), as this is the size of its private cache 
partition.
On the other hand, in a deterministic memory-aware cache,
\emph{bzip2} can allocate more lines from the private partitions of the other 
cores which are not marked as deterministic memory cache lines.
Consequently, the average hit rate is improved by 39\%, 49\%, and 50\%
in DM(A), DM(T98), and DM(T90), respectively, compared with the rate in WP.
Note also that more cache lines are allocated by \emph{bzip2} in 
DM(T98) and DM(T90) compared to DM(A) because more best-effort cache
lines can be available for \emph{bzip2} in these configurations.
The best-effort cache lines of each core's cache partition are shared
among all of the cores, including the core that runs \emph{bzip2} and
those that run the real-time tasks.
We include the result for NoP to show how much cache space \emph{bzip2}
can allocate if there is no restriction.
These numbers can also
be seen as the upper-bound cache space that \emph{bzip2} can allocate in 
the deterministic memory-aware cache. By comparing the cache occupancy
in DM(90) and NoP (i.e., ``free-for-all'' sharing), we see that using
deterministic memory-aware cache, \emph{bzip2}'s cache space
occupancy is close to what we see in NoP.

\subsection{Effects of Deterministic Memory-Aware DRAM Controller} \label{subsec:mem-cntrl}
We evaluate the deterministic memory-aware
DRAM controller, using SD-VBS benchmark suite (input: CIF).
Note that we increase the input size of the SD-VBS benchmarks to
ensure that the working-sets of the benchmarks do not 
fit in the L2 cache and the memory accesses have to go to the main memory.
On the other hand, because the EEMBC benchmarks are cache-fitting and
their working-set size cannot be adjusted, we remove them from this
experiment.
We then re-profile the SD-VBS benchmark with the new
inputs, following the method described in \ref{sec:profile}, to
determine the critical pages.

The basic setup is the same as in \ref{subsec:real-time}: We
schedule a real-time task on Core 0, while co-schedule three
instances of the Bandwidth benchmark as co-runners on Core 1 to 3.
The working-set size of the Bandwidth benchmark is configured to be 2x
larger than the L2 cache size to induce lots of competing DRAM
accesses. We repeat the experiment in the following configurations.
In \emph{DM(A)}, \emph{DM(T98)}, and \emph{DM(T90)}, the cache
configurations are the same as in \ref{sec:result}. In addition, each
core is given a private DRAM bank for deterministic memory in the DM
configurations. The remaining four DRAM banks are shared among the
cores for best-effort memory.
With the DM-aware OS allocator support described in~\ref{sec:dm-dram},
the deterministic memory blocks are allocated on the per-core private
banks, and the best-effort regions are allocated on the shared banks.
In \emph{BA and FR-FCFS}, the FR-FCFS algorithm is
used to schedule the memory accesses to the DRAM, and no OS-level DRAM
bank control is applied (i.e., default buddy allocator).

\begin{figure} [h]
    \centering
    \begin{subfigure}{0.48\textwidth}
        \includegraphics[width=\linewidth]{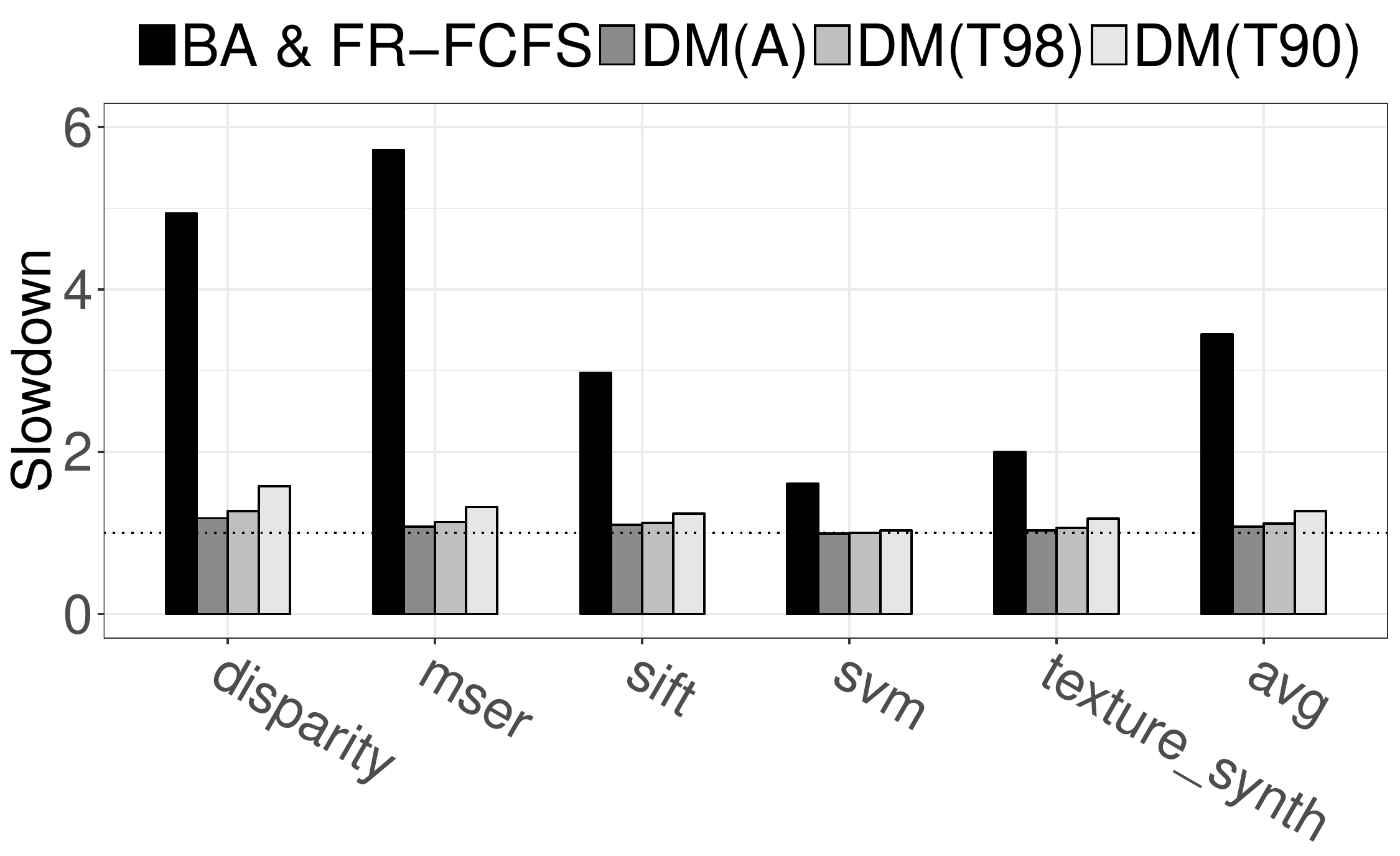}
        \caption{Measured slowdown.}
        \label{fig:slowdown-mc}
    \end{subfigure}
    \hfill
    \begin{subfigure}{0.48\textwidth}
        \includegraphics[width=\linewidth]{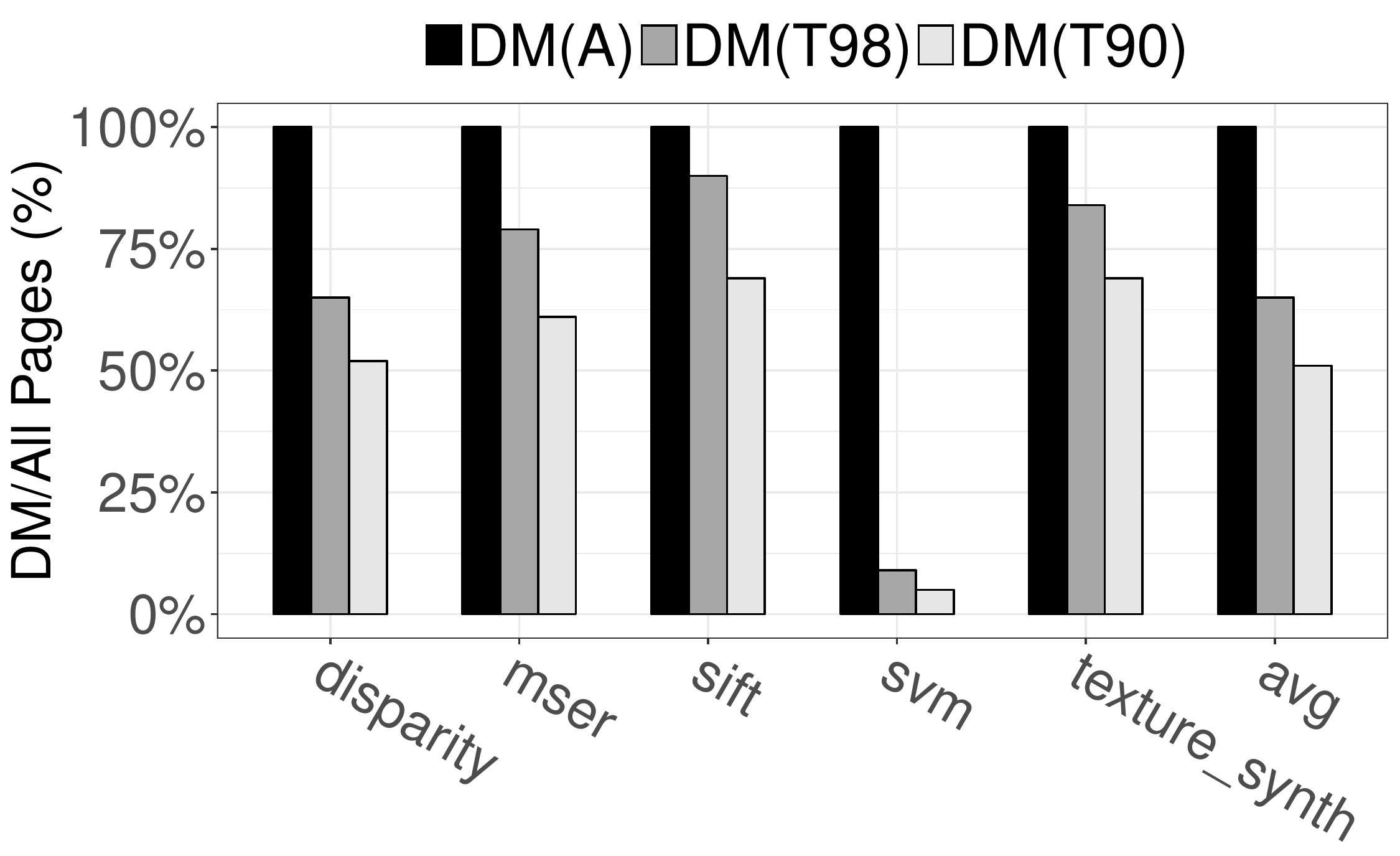}
        \caption{Deterministic memory utilization.}
        \label{fig:dm-util-mc}
    \end{subfigure}

    \caption{Performance and deterministic memory space impacts of
      DM-aware DRAM.}
    \label{fig:eval-mc}
\end{figure}

Figure~\ref{fig:slowdown-mc} shows the normalized slowdown results for
different system configurations. 
Note first that real-time tasks can suffer a significant slowdown in
BA\&FR-FCFS (by up to 5.7X), while all DM-aware configurations suffer
much fewer slowdowns thanks to the two-level scheduling algorithm of
our DM-aware memory controller design.
Figure~\ref{fig:dm-util-mc} shows the ratio between the number pages
marked as deterministic and all the pages touched by each real-time
task. In \emph{DM(A)},
all the pages of each benchmark are marked as deterministic memory,
while in \emph{DM(T90)} only 51\% of pages, on average, are marked as
deterministic 
as more pages are allocated in best-effort DRAM banks.
This space saving is achieved at the cost of slight execution time increase
in real-time benchmarks.
These results show how the number of deterministic pages can be used
as a parameter to make a trade-off between resource utilization and
isolation performance.

\section{Related Work} \label{sec:related}
{\bf Time-predictable hardware architecture.}
Time-predictable hardware architecture has long been studied in the
real-time community.
Edwards and Lee proposed the PRET architecture, which promoted the
idea of making time as a first-class citizen in computer
architecture~\cite{edwards2007pret}. A PRET machine~\cite{liu2012pret} 
provides hardware-based isolation---featuring a thread interleaved
pipeline, scratchpad memory~\cite{banakar2002scratchpad} and a
bank-privatized PRET DRAM controller~\cite{reineke2011pret}---to
support strong timing predictability and repeatability. FlexPRET
improves the efficiency of PRET with a flexible hardware thread scheduler that
guarantees hardware isolation of hard real-time threads while allowing
soft real-time threads to efficiently utilize the processor
pipeline~\cite{zimmer2014flexpret}.
T-CREST~\cite{schoeberl2015t}, MERASA~\cite{merasa2010micro} and
parMERASA~\cite{ungerer2013parmerasa} projects also have investigated
time-predictability focused core architecture, cache, cache coherence
protocol, system-bus, and DRAM controller
designs~\cite{schoeberl2011towards,jordan2013static,rosen2007bus,jalle2014ahrb,paolieri2009hardware,paolieri2009analyzable,li2014dynamic,li2015architecture}.
There are also many other proposals, which focus on improving timing
predictability of each individual shared hardware component---such as 
time predictable shared caches~\cite{yan2010time,yan2011time,lesage2012preti},
hybrid SPM-cache architecture~\cite{zhang2013hybrid}, and
predictable DRAM
controllers~\cite{zheng2013worst,goossen2013conservative,yogen2014ecrts,ecco2015improved}.
In most proposals, the basic approach has been to provide space and
time partitioning of hardware resources to each critical real-time
task or the cores that are designated to execute such
tasks. Thus, CPU-centric abstractions such as task priority and
core/task id are commonly used information sources, which are utilized
by these hardware proposals in managing the hardware resources.
However, when it comes to managing memory related hardware resources,
these CPU-centric abstractions can be too coarse-grained, which make
efficient resource management difficult. This is because neither all
tasks are time-critical (and thus requires hardware isolation
support), nor all memory blocks of a critical task are necessarily
time-critical, as we have shown in Section~\ref{sec:profile}.

{\bf Memory address based real-time architecture designs.}
The basic idea of using physical memory address in hardware-level
resource management has been explored in several prior works.
Kumar et al. proposed a criticality-aware cache design, which uses a
number of hardware range registers to declare critical memory
regions. Its Least Critical (LC) cache replacement algorithm then
prioritize the cache-lines of critical memory regions over others to
ensure predictable cache performance for a single-core, fixed-priority
preemptive scheduled system setup~\cite{kumar2014cache}.
Kim et al. similarly declare critical memory regions using a set of
hardware range registers to distinguish memory-criticality at the
DRAM controller level~\cite{kim2015predictable}.
While our approach is also based on memory address based criticality
determination, our deterministic memory abstraction is designed to be
utilized by the entire memory hierarchy whereas the prior works
focused on a single individual hardware resource
management. Furthermore, a key contribution of our approach is that,
in our approach, memory criticality is determined at the page
granularity by utilizing memory management unit (MMU),
which enables more flexible and fine-grained (page-granularity) memory 
criticality control. In contrast, the prior works may be 
limited by the number of available hardware range registers in
declaring critical memory regions. As such, our MMU-based approach is
compatible with high-performance processors and general-purpose OSs
such as Linux, whereas the prior works primarily focus on MMU-less
processors and RTOSs. We would like to note, however, that our
MMU-based deterministic memory abstraction can be integrated into and
leveraged by these prior works. The deterministic memory abstraction
provides a general framework for the entire memory-hierarchy and thus
is complementary to the prior works.

{\bf OS-level shared resource management.}
In many OS-level resource management approaches, MMU has been a
vital hardware component that the OS leverages for implementing certain
memory management policies for real-time systems. Page-coloring is a
prime example that has been used to partition shared
cache~\cite{liedtke97ospart,lin2008gaining,zhang2009towards,soares2008reducing,ding2011srm,ward2013ecrts,mancuso2013rtas,kim2013coordinated},
DRAM banks~\cite{yun2014rtas,liu2012software,suzuki2013coordinated}
and even TLB~\cite{panchamukhi2015providing} by selecting certain
physical addresses (cache color, DRAM bank, etc.) in allocating pages.
However, in most OS-level resource management approaches, shared
resources are allocated at the granularity of task or core, which is
too coarse-grained and therefore can result in resource under-utilization
problems. Furthermore, these OS-level 
resource management approaches have fundamental limitations because
they generally cannot directly influence important resource allocation
and scheduling decisions done by the underlying hardware due to the
lack of a generalized abstraction that allows such cross-layer
communication. We address these limitations by proposing the
deterministic memory abstraction, which enables close collaboration
between the OS and the underlying hardware components in the memory
hierarchy to achieve efficient and predictable resource allocation and
scheduling.
To the best of our knowledge, we are first to propose to encode each
individual memory page's time criticality in the page's page table
entry, which is then passed through the entire memory hierarchy to
enable system-wide, end-to-end memory-criticality-aware resource
management.

\section{Conclusion and Future Work}\label{sec:conclusion}

In this paper, we proposed a new memory abstraction, which we call
\emph{Deterministic Memory}, for predictable and efficient resource
management in multicore. We define deterministic memory as a special
memory space where the platform---OS and hardware
architecture---guarantees small and tightly bounded worst-case access
timing.

We presented OS and architecture extensions to efficiently support the
deterministic memory abstraction. In particular, we presented a
deterministic memory-aware cache design that leverages the abstraction
to improve the efficiency of shared cache without losing isolation
benefits of traditional way-based cache partitioning.
In addition, we proposed a deterministic memory-aware DRAM
controller which effectively reduces the necessary core-private DRAM
bank space while still providing good isolation performance.
We implemented the proposed OS extension on a real operating system  
(Linux) and implemented the proposed architecture extensions on a
cycle-accurate full-system simulator (gem5). 

Evaluation results show the feasibility and effectiveness of
deterministic memory based cross-layer resource management.
Concretely, by using deterministic memory, we achieved the same degree
of strong isolation while using 49\% less cache space, on average,
than the conventional way-based cache partitioning method.
Similarly, we were able to reduce required private DRAM bank space
while achieving comparable isolation performance for DRAM intensive
real-time applications, compared to a baseline real-time DRAM controller.

We are currently working on implementing the proposed architecture
extensions on a FPGA using an open-source RISC-V based multicore
platform~\cite{boom-github}.
We also plan to develop methodologies and tools to identify
``optimal'' deterministic memory blocks that maximize the overall
schedulability.



\section*{Acknowledgments} \label{acknowledge}
This research is supported by NSF CNS 1718880. Any opinions, findings,
and conclusions or recommendations expressed in this publication are
those of the authors and do not necessarily reflect the views of the
NSF.



\bibliography{related}

\end{document}